\documentclass{aa}  

\usepackage{graphicx}
\usepackage[dvipsnames]{xcolor}
\usepackage{caption}
\usepackage{multirow}
\usepackage{txfonts}
\usepackage[]{hyperref}

\hypersetup{colorlinks,citecolor=blue}
\usepackage{threeparttable}
\usepackage{epstopdf}

\begin{document} 

   \title{Spectral state transitions in Circinus ULX5}


   \author{Samaresh Mondal\inst{1}\thanks{E-mail: smondal@camk.edu.pl (SM)}, 
   		   Agata R{\'o}{\.z}a{\'n}ska\inst{1}
   		   \and 
   		   Patrycja Bagi{\'n}ska\inst{2}
   		   \and 
   		   Alex Markowitz\inst{1,3}
   		   \and 
   		   Barbara De Marco\inst{4}
          }

   \institute{Nicolaus Copernicus Astronomical Center, Polish Academy of Sciences,
              ul. Bartycka 18, 00-716 Warsaw, Poland
         \and
         Astronomical Observatory Institute, Faculty of Physics, A.~Mickiewicz University, S{\l}oneczna 36, 60-286 Pozna\'n, Poland
         \and
         University of California, San Diego, Center for Astrophysics and Space Sciences, MC 0424, La Jolla, CA, 92093-0424, USA
		\and
         Departament de Física, EEBE, Universitat Politècnica de Catalunya, Av. Eduard Maristany 16, E-08019, Barcelona, Spain
             }

   \date{Received XXX; accepted YYY}
	
   \authorrunning{Mondal et al.}
   \titlerunning{Circinus ULX5}
   
 
  \abstract
   {We performed timing and spectral analysis of multi-epoch {\it Suzaku}, {\it XMM-Newton} 
   and {\it NuSTAR} observations of the ultraluminous X-ray source (ULX) Circinus ULX5, to put
   constraints on the mass of the central object and the accretion mode operating in this source.}
   {We aim to answer whether the source contains a stellar mass black hole with a super-Eddington 
   accretion flow or an intermediate mass black hole accreting matter in a sub-Eddington mode. Moreover, 
   we search for major observed changes in spectra and timing occur, and determine if they are associated with major 
   structural changes in the disk, similar to those in black hole X-ray binaries.}
   {We collected all available broadband data from 2001 to 2018 including \emph{Suzaku}, 
   \emph{XMM-Newton} and \emph{NuSTAR}. We performed timing and spectral analysis to study the relation 
   between luminosity and inner disk temperature. We proceeded with time-averaged spectral analysis 
   using phenomenological models of different accretion modes. Finally, we constructed the hardness 
   ratio versus intensity diagram to reveal spectral state transitions in Circinus ULX5.}
   {Our spectral analysis revealed at least three distinctive spectral states of Circinus ULX5, in analogy 
   to state transitions in  Galactic black hole X-ray binaries. Disk-dominated spectra are found 
   in high flux states and the power-law dominated spectra are found in lower flux states. The 
   source was also observed in an intermediate state, where the flux was low, but spectrum is dominated by a disk 
   component. Over eighteen years of collected data, ULX5 appeared two times in the high, three times in the low, 
   and two times in the intermediate state. The fastest observed transition was $\sim7$
    months.}
   {Our analysis suggests that the central object in Circinus ULX5 is a stellar mass BH (<10\,M$_{\odot}$), 
   or possibly a NS even though we do not detect pulsations in the lightcurves. Fractional variability amplitudes 
   are consistent with state transitions in Circinus ULX5 wherein higher variability from the power law-like Comptonized 
   emission gets suppressed in the thermal disk-dominated state.} 

   \keywords{Accretion disks, Stars: black holes, X-rays: binaries, X-rays: Circinus ULX5}

   \maketitle
%

\section{Introduction}

The biggest mystery of ultraluminous X-ray sources (ULXs) is the mass
of the central compact object.  ULXs are extragalactic
compact accreting objects, located outside the nuclei of galaxies. Their
luminosities typically exceed the Eddington luminosity of a typical
10\,M$_{\odot}$ black hole (BH) accretor. They have high values of X-ray
brightness, $L_{\rm x}$>$10^{39}\ \rm erg\ s^{-1}$, and low inner disk
temperature, $kT_{\rm in}\sim$0.1--0.3 keV, obtained from spectral
fitting of multicolor accretion disk (MCD) components. Consequently,
ULXs have been suggested as prime candidates to host intermediate-mass
BHs (IMBHs), $\sim 10^{2-4}$\,M$_{\odot}$
\citep{kaaret2003,miller2004a,miller2004b}. However, the notion of
every ULX harbouring an IMBH conflicts with population studies, since
it would imply too high an IMBH formation rate in star-forming galaxies
\citep{king2004}. Such a notion is also inconsistent with the break at $\sim
2\times 10^{40}$\,erg\,s$^{-1}$ in the luminosity function of
point-like X-ray sources in star-forming galaxies
\citep{swartz2004,mineo2012}.

A breakthrough discovery was made in 2014, when a pulsation with an
average period of 1.3 s was measured in the \emph{NuSTAR} data of M82
X-2 \citep{bachetti14}. Several subsequent discoveries of ULX pulsars
\citep{furst2016,israel2017a,israel2017b,brightman2018,carpano2018,sathyaprakash2019,rodriguez-castillo2020},
indicated that some ULXs have a neutron star (NS) as an accretor. The
optical observations of a limited number of ULXs' companions revealed
that they might belong to the class of High-Mass X-ray Binaries
\citep{motch2011,motch2014,heida2015,heida2016}. These objects
may comprise some fraction of the merging double compact objects detected by advance LIGO
and Virgo \citep{mondal2020}.

Due to the lack of a method for direct measurement of the compact
object's mass in ULXs, we can rely only on indirect procedures, e.g.,
by inferring the accretion mode of the source and comparing it to
those of known sources.  The most promising and easy test is to study
the broadband spectra of ULXs, with the aim of deriving correlations
between measurable global parameters. It is well known that in the
case of Galactic black hole X-ray binaries (XRBs), if an observer has a
direct view on the hot inner accretion disk, the standard model by
\citet[][; hereafter SS73]{ss73} predicts a relation between
luminosity and temperature: $L_{\rm disk}\propto T_{\rm in}^4$
\citep[measured e.g.\ by][]{gierlinski2004}.  However, when the object is viewed close to edge-on, the
source is visible through a disk wind photosphere that is measured to
have a temperature of 0.1--0.3 keV. 
 The lower value of temperature results from the fact that when the
 source luminosity increases, the size of the photosphere becomes
 larger and its temperature drops. As demonstrated by
     \cite{king2002} and \cite{king2009}, this leads to an inverse
     luminosity relation in sources accreting at super-Eddington
     rates; \citet{king2016} demonstrated that such thick disk winds
     can be produced in super-Eddington accreting ULXs.
This inverse correlation between disk luminosity and temperature has been observed 
in many individual ULXs
after fitting phenomenological models as MCD plus a power law to their broadband X-ray spectra   
\citep{feng2007,kajava2009,mondal2020b}. Evidence for disk winds was also found in several 
ULXs via detection of emission and absorption lines in high-resolution spectra 
\citep{middleton2014,pinto2016,walton2016,pinto2017,kosec2018a,kosec2018b}. 
Thus, the combination of an inverse disk luminosity -- temperature relation and detection of a disk wind can 
support the notion that a given ULX source harbors a stellar-mass BH with super-Eddington accretion. 

Another indirect constrain on BH mass in ULXs comes from X-ray spectral variability. 
It was found that some ULXs can represent different spectral states 
\citep[for review see:][]{kaaret2017} in analogy with those observed in XRBs. 
Across the known population of ULXs, we can distinguish the following
spectral states depending on X-ray luminosity: 

\begin{itemize}
\item At $L_{\rm X} > 3\times10^{39}$\,erg s$^{-1}$, sources display a
so-called hard or soft UL state.  The spectrum is fitted by two
components, a soft excess and a hard component which shows a turnover
at high energies. In the soft UL state, the soft component dominates
the luminosity \citep{sutton2013}.

\item At $L_{\rm X} < 3 \times 10^{39}$\,erg~s$^{-1}$, one finds the broadened disk (BD) state:  A standard MCD model component
is too narrow to fit the spectrum, and so it has to be broadened by a slim disk model or power-law component \citep{sutton2013}.

\item Finally, thermal spectrum with color temperature $kT \sim 0.1$\,keV and bolometric luminosity 
$\sim$ a few times $10^{39}$\,erg~s$^{-1}$ with almost no emission above 1 keV are classified as supersoft UL 
(SSUL) state: Spectra are dominated by a low temperature blackbody component with  which produces
over 90\% of the intrinsic flux in the 0.3--10 keV band \citep{kong2003}. \end{itemize}

Some of the highest quality {\it XMM-Newton} spectra of a few tens of
ULXs display a connection between spectral state and accretion mode
\citep{roberts07,gladstone2009,sutton2013}: the BD state is usually
observed in sources with lower luminosities, suggesting sub- or nearly
Eddington accretion, while two-component UL states are seen in
high-luminosity sources that strongly support super-Eddington
accretion.
Furthermore, the majority of BD state objects do not display strong
short timescale variability, and the fact that only a few ULXs are strongly variable may be evidence that in those
sources, a two-component hard/soft UL spectrum is present.
The UL state objects differ in their variability properties, such that
hard UL spectra display lower levels of fractional variability ($<<
10$\%) while soft UL spectra are highly variable (10--30\%). The
difference in variability properties may be due to the wind opening
angle, where geometrical beaming occurs, leading to eventual
obscuration of hard radiation by the thick wind \citep{sutton2013}.
However, it is important to note that this behaviour is opposite to
that seen in XRBs, where sources in the thermal-dominated spectral
state exhibit lower variability than in the power law-dominated hard
state \citep{churazov2001,munoz2011}.  Thus, the detection of a
spectral state transition while tracking corresponding short time
scale variability may indicate the accretion mode, and consequently,
the BH mass in ULXs.

However, actual state transitions within individual sources are poorly studied, since 
multi-epoch observations are needed. Up to now, a single transition from soft-to-hard 
and hard-to-soft state was reported in the cases of IC 342 X-1 and IC 342 X-2, respectively 
\citep{kubota2001}, and in the case of NGC 1313 X-1 \citep{feng2006a}. Furthermore, the brightest ULX 
source in NGC~274 displayed a state transition between soft UL and SSUL, the latter representing a
higher accretion rate \citep{feng2016,pinto2021}.  For extreme super-Eddington accretion, blackbody emission may arise from 
the photosphere of a thick outflow and hard X-ray emission only emerges from the central low-density funnel. 
Until now, only Holmberg IX X-1 exhibited three spectral states, over a span of eight years \citep{luangtip2016}, and 
most probably due to enhanced geometric beaming as the accretion rate increases and the wind funnel 
narrows; such changes cause the scattered flux from the central regions of the super-Eddington flow to brighten 
faster than the isotropic thermal emission from the wind. Thus the detection of a state transition within an 
individual source may put constraints on the possible geometry of the emitting compact object. 

In this paper, we aim to determine the nature of the compact object
Circinus ULX5 and constrain its mass, as well as explore its accretion flow 
properties as a function of luminosity. To this end, we present timing and spectral analysis of broad-band 
X-ray data of Circinus ULX5 from 
different epochs during 2001--2018 taken by {\it Suzaku} \citep{mitsuda2007}, {\it XMM-Newton} 
\citep{jansen2001} and {\it NuSTAR} \citep{harrison2013a}. Our aim is to put tight constraints 
on the BH mass using all indirect methods described above. 
The paper is organized as follows: in $\S$~\ref{sec:obs} we describe the source and summarize its current 
status known from X-ray data. In the same section, we present the observations used in this paper and the data 
reduction processes. The next three sections contain results obtained from short timescale variability, the hardness-intensity diagram 
(HID), and broadband spectral analysis, respectively. We discuss our results in $\S$~\ref{sec:disc}, and put conclusions 
in $\S$~\ref{sec:concl}.

\section{Observations and  data reduction}
\label{sec:obs}
	
	\begin{table*}
		\begin{center}
		\caption{The details of X-ray data analyzed in this paper. Columns from the left display: 
		the name of the data set, observation ID, the date, exposure time, good time intervals (GTI) and net 
		total counts for different detectors, i.e. for \emph{XMM-Newton} pn/MOS1/MOS2, for \emph{Suzaku} 
		XIS0+2+3(added)/XIS1, and for \emph{NuSTAR} FPMA/FPMB. 
		The total counts are given for the following energy ranges: 
                0.5--10 keV for \emph{XMM-Newton} and \emph{Suzaku}, and 3--30 keV for \emph{NuSTAR}.
		The unabsorbed flux has been computed using a model composed of  \texttt{diskbb+pl} 
		and the luminosity was calculated assuming a distance of 4.2\,Mpc. References to the published data are 
		shown in the last column.
		[1] = First detection of the source by \citet{2006-Winter}; 		
		[2] = First joint \textit{XMM-Newton} + \textit{NuSTAR} observation by \citet{walton13}; 		
		[3] = Data fitted by single component model by \citet{rozanska2018}. 
                }
		\setlength{\tabcolsep}{3pt}               	
    		\renewcommand{\arraystretch}{1.5}			
			\begin{tabular}{lrccccccc}
			\hline\hline
			\multirow{2}{*}{Data set} & \multirow{2}{*}{ObsID} & \multirow{2}{*}{Date} & \multirow{2}{*}{$T_{\rm exp}$ (ks)} & \multirow{2}{*}{GTI (ks)}
			& \multirow{2}{*}{Total counts} & $F_{\rm 0.5-10 keV}$ & $L_{\rm 0.5-10 keV}$  & \multirow{2}{*}{Ref.}\\
			&&&&&& [$\rm erg\ s^{-1}\ cm^{-2}$] & [$\rm erg\ s^{-1}$]\\ \hline
			\emph{XMM-01} & 0111240101 & 2001-08-06 & 109.8 & 70.5/98.2/- & 8563/10977/- & $2.11\times 10^{-12}$ & $4.45\times 10^{39}$ & [1] \\
			\hline
			\emph{Suzaku-06} & 701036010 & 2006-07-21 & 108 & 108/108 & 27220/7433 & $4.19 \times 10^{-12}$ & $8.45 \times 10^{39}$ & [2] \\
			
			\hline 
			\emph{XMM-13} & 0701981001 & 2013-02-03 & 58.9 & 37.2/45.6/48.1 & 39565/15149/18212 & \multirow{2}{*}{$8.52\times10^{-12}$} & 
			\multirow{2}{*}{$1.80\times10^{40}$} & \multirow{2}{*}{[2],[3]}\\
			\emph{NuSTAR-13} & 30002038004 & 2013-02-03 & 40.3 & 40.27/40.21 & 4353/4179 & & &   \\
			\hline
			\emph{XMM-14} & 0656580601 & 2014-03-01 & 45.9 & 23.8/30.4/30.2 & 5888/2642/1909 & $3.46\times 10^{-12}$ & $7.30\times 10^{39}$ & -- \\
			\hline
			\emph{XMM-16} & 0792382701 & 2016-08-23 & 37.0 & 28.6/5.9/5.0 & 9712/834/946 & \multirow{2}{*}{$3.65\times10^{-12}$} & \multirow{2}{*}{$7.70\times10^{39}$}  &  \multirow{2}{*}{--} \\
			\emph{NuSTAR-16} & 90201034002 & 2016-08-23 & 49.8 & 49.8/49.7 & 579/750 &  \\
			\hline
			\emph{XMM-18F} & 0780950201 & 2018-02-07 & 45.7 & 5.7/12.7/20.0 & 929/863/1413 & $2.81\times10^{-12}$ & $5.93\times10^{39}$ & --\\
			\hline 
			\emph{XMM-18S} & 0824450301 & 2018-09-16 & 136.2 & 86.21/114.9/112.4 & 90239/34360/38045 & $7.60\times10^{-12}$ & $1.60\times10^{40}$ & -- \\
			\hline
			\label{tab:obs}
		\end{tabular}
		\end{center}
	\end{table*}
	
Circinus ULX5 is located on the outskirts of the Circinus galaxy, at $\rm \alpha=14^h12^m39^s$ and 
$\rm \delta=-65^{\circ}23^{'}34^{''}$. The source was first detected
in an \textit{XMM-Newton} observation taken in 2001, and reported the ULX catalog of
\citet{2006-Winter} under the name Circinus XMM2.
The source was interpreted to be in the so-called 
``high-state'', in which the thermal component dominates. 

The first observations of ULX5 by the {\it NuSTAR} telescope were performed and analyzed by 
\citet{walton13}. While the data taken in 2001 and 2006 by {\it XMM-Newton} indicated extreme 
flux variability, the later observation in 2013 did not show signatures of variability. Based 
on {\it Suzaku} observations in 2006 \citep{walton13}, ULX5 displays long term spectral evolution 
based on a clear correlation of hardness ratio with 0.5--10 keV luminosity. The increasing positive 
trend in the hardness-luminosity diagram is accompanied by 0.5--10 keV fractional variability 
amplitude reaching $\sim 12$\%. The source peak luminosity was calculated to reach a value of 
$1.6 \times 10^{40}$ erg s$^{-1}$ for a distance of $\sim 4$\,Mpc \citep{walton13}, measured with 
the Tully--Fisher method \citep{1977-Tully} when the Circinus galaxy ($z$ = 0.001448) was discovered 
\citep{freeman77}. In X-ray spectral fits,
depending on the model used, the values of Galactic neutral absorption toward 
the source range from 1 to $9.5 \times 10^{21}$~cm$^{-2}$. These values are consistent with the reported column densities
$N_{\rm H I}  + N_{\rm H2}$ $\sim$ (5.01+1.44)$\times10^{21}$\,cm$^{-2}$ of Galactic monatomic and diatomic hydrogen, respectively, toward the 
source by \citet{willingale2013} and \citet{bekhti2016}.

The details of X-ray observations we studied are given in Tab.~\ref{tab:obs}.
Henceforth in this paper we use the names of data sets listed in the first column 
of Tab.~\ref{tab:obs}. Four data sets taken from 2014 to 2018 have been never published before.  
When calculating the luminosity, we accept the distance of 4.2\,Mpc to the Circinus galaxy as discussed in 
Appendix~\ref{sec:dist}. We also studied the RGS (reflecting grating spectrometer) high resolution spectra 
taken in 2013, 2014, and 2018, but the data did not show any strong evidence for a disk wind; these data are presented 
in Appendix~\ref{sec:rgs}. 

\subsection{XMM-Newton}

We analyzed six \emph{XMM-Newton} observations performed between 2001 to 2018, four of them never 
published before. The data reduction was done using XMM-Newton Science Analysis System (SASv16.0.0) 
following the online procedures. The first-order observation data files (ODF) were processed using 
\texttt{emproc} and \texttt{epproc} to create clean calibrated event files for the EPIC-MOS and EPIC-pn 
detectors, respectively. The source products were extracted from a circular region of $40\arcsec$ radius 
and the background was selected from a source free region of size $80\arcsec$. We used \texttt{evselect} 
to create lightcurves and spectra. We allowed only single and double events for the EPIC-pn detector and single 
to quadruple events for both EPIC-MOS detectors. The source lightcurves were corrected for background 
counts using \texttt{epiclccorr}. We used  \texttt{rmfgen} and \texttt{arfgen} to generate the redistribution matrix and auxiliary response
files. We note here, that in some of the \emph{XMM-Newton} observations the 
EPIC-pn and MOS detectors were not observing simultaneously. This resulted in different good time intervals (GTI) 
for the three detectors. Unfortunately, in case of \emph{XMM-01}, observations with MOS-2
were taken in "partial window" mode; the source fell onto the outer part of the CCD and was read out, so no data on the ULX were collected.

\subsection{Suzaku} 
\label{obs:suz}
We used \emph{Suzaku} data taken in 2006 using the XIS (X-ray Imaging Spectrometer) CCDs (see Tab.~\ref{tab:obs},
second row). Data reduction was performed using the HEASOFT software package.  For spectral analysis 
we have used cleaned event files from the front-illuminated CCDs --- XIS0, XIS2 and XIS3 --- and back-illuminated CCD ---XIS1, 
for both available editing modes, 3x3 and 5x5. We used \texttt{xselect} to extract lightcurves and spectra. 
The source and background products were extracted from a circular region of 85$\arcsec$ in radius. The background 
was derived from regions free of other sources but in close 
proximity to the ULX. The attitude correction was performed with the use of the \texttt{aeattcor2} standard FTOOL. We 
used the \texttt{xisresp} script to generate response matrix files,
and the \texttt{xisarfgen} script to create auxiliary response files, 
for each detector and editing mode separately. 
We added all front-illuminated CCDs, and therefore the total counts reported in Tab.~\ref{tab:obs}
are combined from all three chips.
 
These data were published previously by \citep{walton13}, and the lightcurve presented in Fig.~9 of their paper 
clearly indicates a drop in flux by a factor of two, at around 40 ks into the observation. Therefore, 
for the purpose of this paper, we extract {\it Suzaku} lightcurves and spectra from the first and second epochs of 
the observations: {\it Suz-H-06} spanned the first $40.08$~ks of the exposure while 
{\it Suz-L-06} spanned the last $67.94$~ks. For both epochs' spectra, we created combined 
spectra for the front-illuminated detectors (XIS023), using the FTOOL \texttt{addascaspec}.
Furthermore, for both XIS023 and back-illuminated XIS1 spectra we combined 3x3 and 5x5 modes using the same FTOOL. 
The process of combining and subtracting lightcurves was performed using the XRONOS script \texttt{lcmath}. 

\subsection{NuSTAR} 
The field of the Circinus galaxy was observed by \emph{NuSTAR} in 2013 and in 2016. The data reduction was done 
using the pipeline \emph{NuSTAR} Data Analysis Software (NUSTARDAS v1.9.5 provided with HEASOFT). The unfiltered event files were 
cleaned and data taken during passages through the South Atlantic Anomaly were removed using \texttt{nupipeline}. The source products 
were extracted from a circular region of radius $70\arcsec$ using \texttt{nuproducts}. The background was selected 
from a source free circular region of radius $100\arcsec$.

	
\section{Short timescale variability} 

	\begin{figure*}[]
    	\includegraphics[trim={0.5cm 0 0 0},width=0.24\textwidth, angle=270]{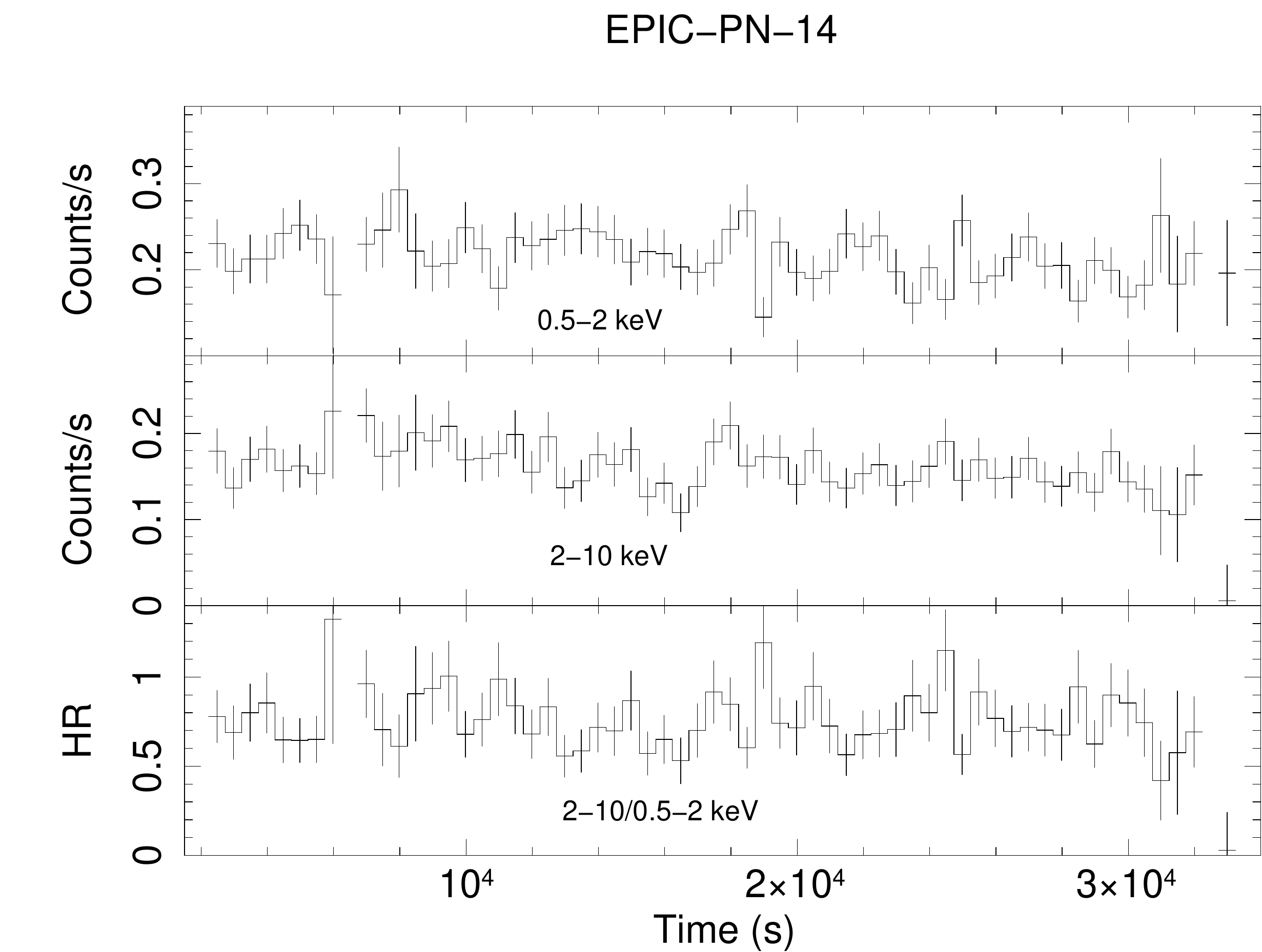}
    	\includegraphics[trim={0.5cm 0 0 0},width=0.24\textwidth, angle=270]{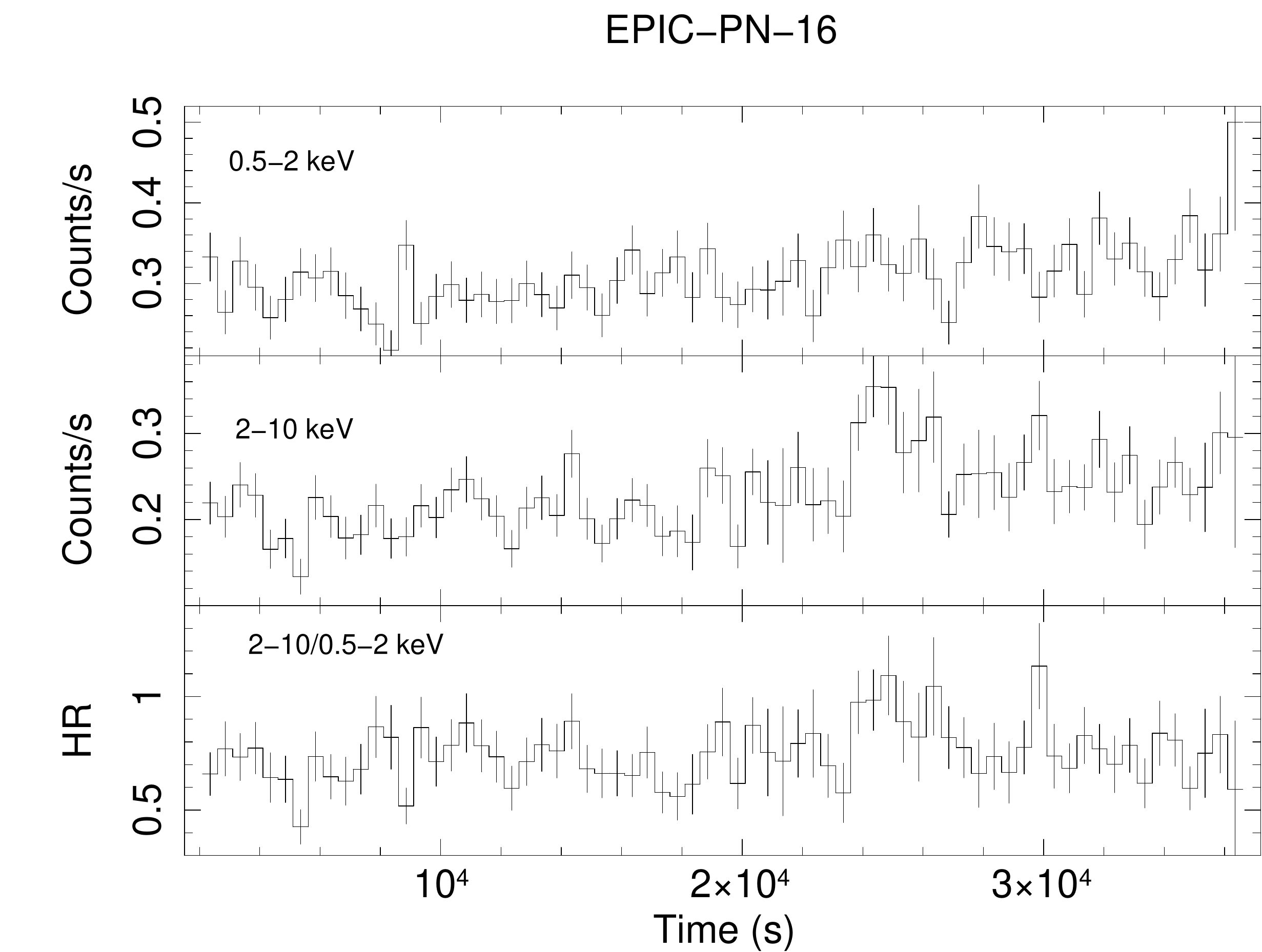}
    	\includegraphics[trim={0.5cm 0 0 0},width=0.24\textwidth, angle=270]{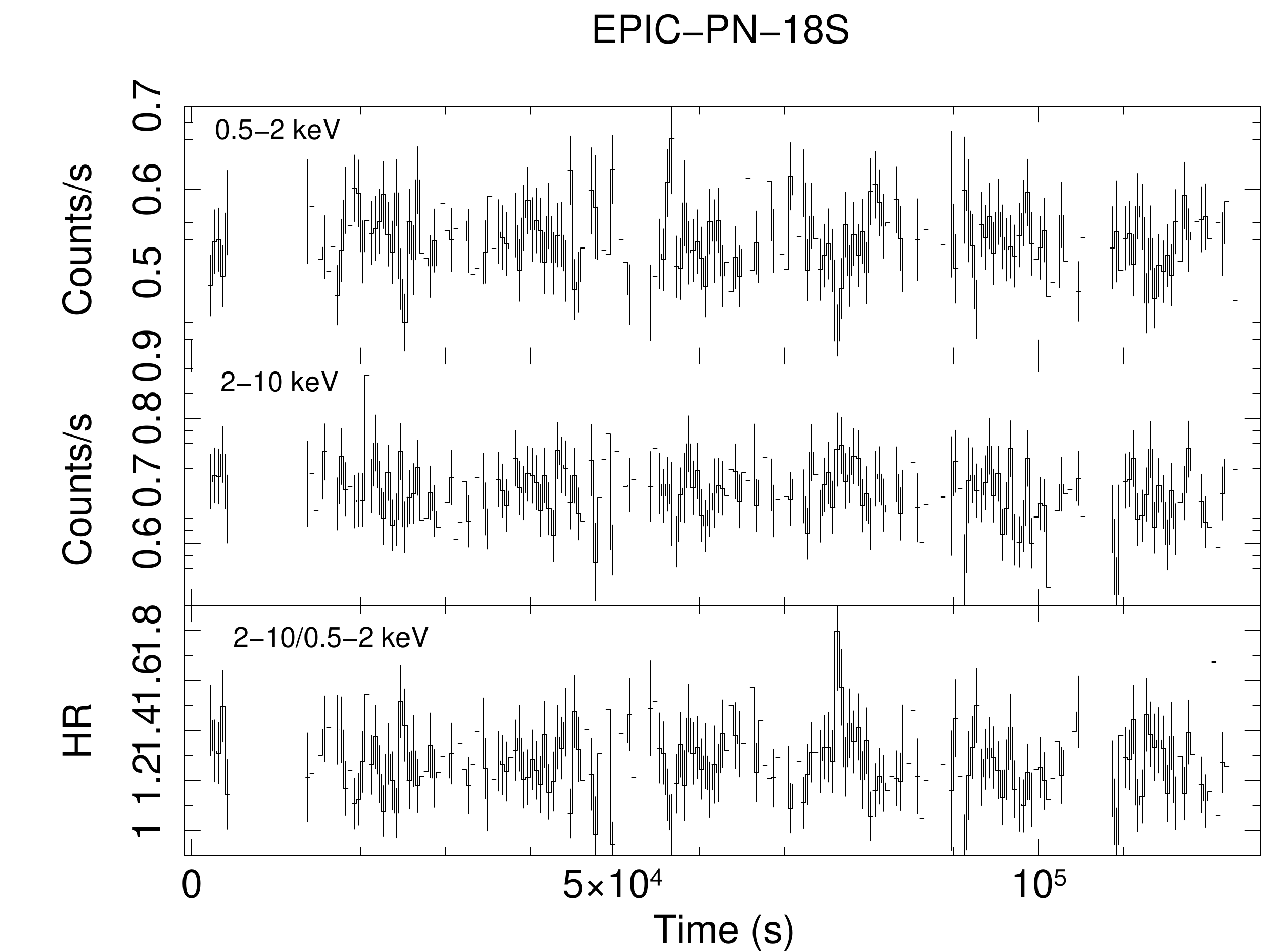}
    	\caption{The lightcurves extracted from the EPIC-pn detector, from data set \emph{XMM-14} (left panel), 
    	\emph{XMM-16} (middle panel), and \emph{XMM-18S} (right panel). In each case, we present lightcurves for 
    	the soft (0.5--2 keV) and hard (2--10) keV bands (upper and middle panels, respectively) as well as the 
    	2--10/0.5--2 keV hardness ratio (bottom panels). The lightcurves were re-binned to 500 s 
    	to yield a higher signal to noise ratio within each bin.}
    	\label{fig:xmmlight}
	\end{figure*}
	
Circinus ULX5 is an extremely variable ULX compared to other ULXs of similar luminosity, which 
typically do not show strong flux variability \citep{heil2009}. \cite{walton13} found that ULX5 
showed 0.5--10 keV fractional rms variability amplitudes
\citep[hereafter $F_{\rm var}$;][see in the last paper Eqs.~10 and B2 for variance and variance's error]{edelson2002,vaughan2003}. 
of $\sim$12--15$\%$ during the 2001 
\emph{XMM-Newton} and 2006 \emph{Suzaku} observations. In the 2013 \emph{XMM-Newton} observation, however, 
$F_{\rm var}$ dropped to $<2\%$. 

We start our analysis with lightcurves. The average count rates in the EPIC-pn detector in 
the 0.5--10 keV band are 0.38, 0.54 and 1.22 ct s$^{-1}$ for \emph{XMM-14}, \emph{XMM-16} and \emph{XMM-18S}, 
respectively. All lightcurves are re-binned to 500~s. Fig.~\ref{fig:xmmlight} shows the 
soft (0.5--2 keV) and hard (2--10 keV) lightcurves as well as the hardness ratio (2--10 keV/0.5--2 keV) 
for three unpublished data sets.

Then for each epoch, we calculate $F_{\rm var}$ in three spectral bands, 0.5--10, 0.5--2, and 2--10 keV.
For longer observations, \emph{XMM-01}, \emph{Suz-H-06}, \emph{Suz-L-06} and \emph{XMM-18S}, we chopped the 
lightcurves into 30 ks segments to ensure a more systematic comparison
to the other observations. $F_{\rm var}$ was calculated 
from each segment separately and then averaged. The lightcurves have a time binning of 500 s, so $F_{\rm var}$ 
gives the power spectrum integrated over the frequency range $3.33\times10^{-5}$ Hz (1/30ks) to $10^{-3}$ Hz (1/$2\times500$s). 
The variability amplitudes calculated for the total, soft, and hard bands for all eight data sets are 
listed in Tab.~\ref{tab:var}. The source shows 0.5--10 keV fractional excess variance at the 
$\sim$10-15\% level during  \emph{XMM-01}, \emph{Suz-H-06}, \emph{Suz-L-06}, \emph{XMM-14}, \emph{XMM-16} and 
\emph{XMM-18F} observation, whereas the value has dropped to $\sim$2\% in \emph{XMM-13} and \emph{XMM-18S} 
data, consistent with earlier studies. In the case of the \emph{XMM-18F} lightcurves,
only an upper limit on $F_{\rm var}$ could be measured or else $F_{\rm var}$ was not constrained.

	\begin{table}
		\centering
		\caption{The fractional variability amplitude $F_{\rm var}$ obtained from the total (0.5--10 keV), soft (0.5--2 keV), and hard 
		  (2--10 keV) band lightcurves, for the data sets given in the first column. A dash indicates that
                  variability amplitude could not be constrained.}
		\setlength{\tabcolsep}{3pt}               	
    		\renewcommand{\arraystretch}{1.5}			
		\begin{tabular}{cccc}
			\hline\hline
			\multirow{3}{*}{Data set} & $F_{\rm var}$ & $F_{\rm var}$ & $F_{\rm var}$\\
			& 0.5-10 keV & 0.5-2 keV & 2-10 keV\\
			& (\%) & (\%) & (\%)\\ \hline

			\emph{XMM-01} & $9.38\pm6.70$ & $6.11\pm15.86$ & $12.76\pm7.95$\\
			\emph{Suz-H-06} & $12.78\pm1.07$ & $8.25\pm2.34$ & $15.97\pm1.37$\\
			\emph{Suz-L-06} & $10.08\pm2.34$ & $8.61\pm4.67$ & $13.03\pm3.23$\\
			\emph{XMM-13} & $2.80\pm1.09$ & $1.91\pm3.12$ & $3.16\pm1.61$\\
			\emph{XMM-14} & $14.07\pm1.76$ & $11.45\pm2.65$ & $17.04\pm3.04$\\
			\emph{XMM-16} & $9.79\pm1.28$ & $5.09\pm2.26$ & $14.23\pm2.28$\\
			\emph{XMM-18F} & $< 6.22$ & -- & $<11.18$\\
			\emph{XMM-18S} & $1.33\pm3.52$ & $3.32\pm3.41$ & $1.51\pm5.63$\\
			\hline
		\end{tabular}
		\label{tab:var}
	\end{table}
	
\section{Hardness-intensity diagram}
\begin{figure}[]
    	\centering
    	\includegraphics[width=0.48\textwidth]{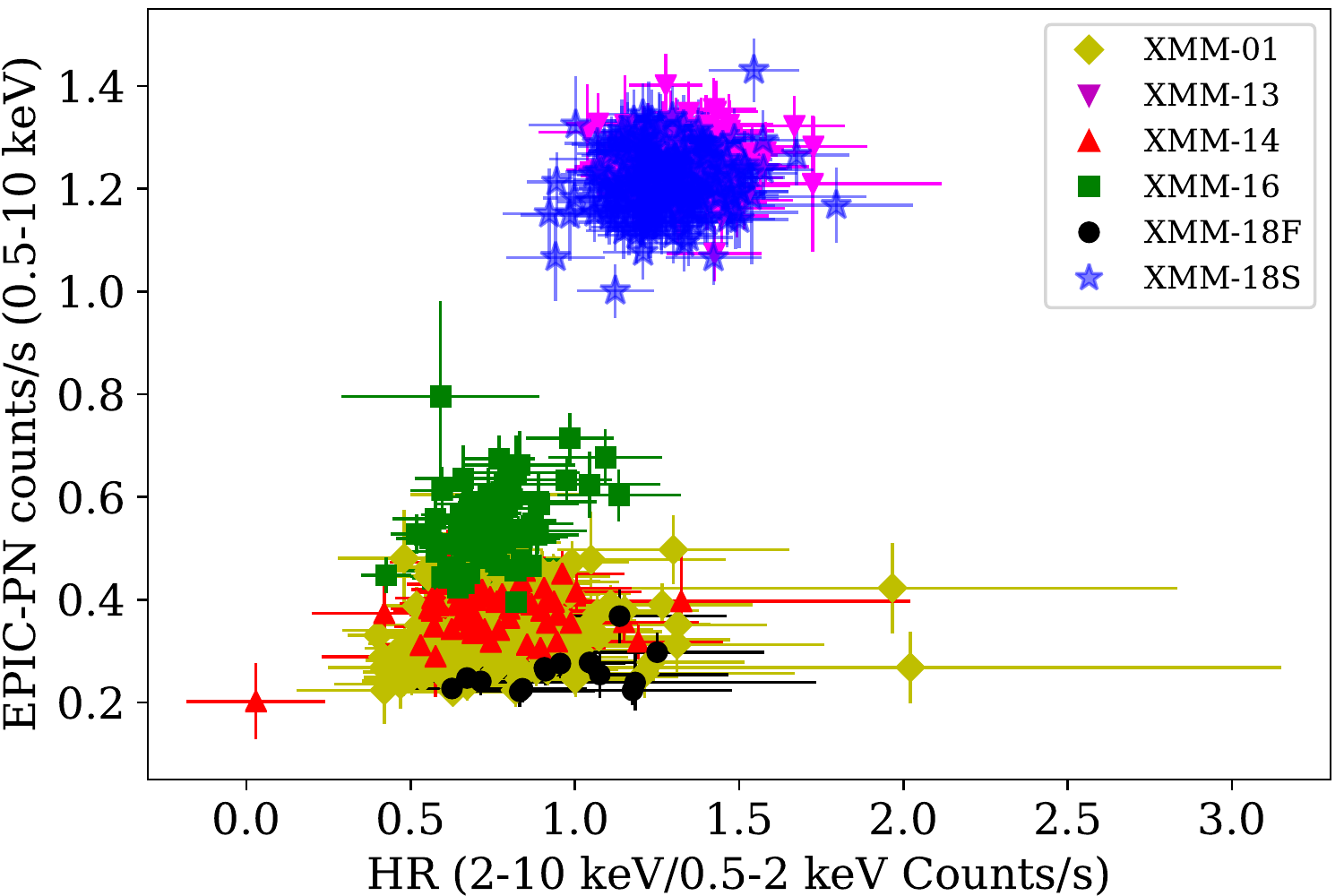}
    	\caption{The hardness ratio intensity diagram made from the EPIC-pn lightcurves. 
    	All data sets are arranged in two islands. }
    	\label{fig:hr_count}
	\end{figure}

In order to search for potential spectral state transitions, we plotted the count rate hardness ratio versus 
intensity diagram using the EPIC-pn lightcurves from all data sets.
We again used lightcurves binned to 500~s, and again using 
0.5--2, 2--10, and 0.5--10 keV. Fig.~\ref{fig:hr_count} shows 
the (2--10 keV)/(0.5--2 keV) count rate ratio (hardness) plotted 
against the total band 0.5--10\,keV count rate (intensity).
There are two distinct ``islands'' visible in 
the diagram, where the intensity differs by a factor of 4--5, and 
the hardness ratio in the extreme cases differs by 1. Furthermore, 
the \emph{XMM-16} data (green squares in Fig. \ref{fig:hr_count}) 
are a factor of two brighter than  \emph{XMM-01}, \emph{XMM-14} 
and \emph{XMM-18F} at the same hardness ratio. \emph{Suzaku} data are 
not included in this diagram since the difference between count rate of both 
instruments does not allow for a direct comparison; nevertheless the HID for the 2006 \textit{Suzaku} data was published 
by \citet{walton13}.

It is evident from Fig.~\ref{fig:hr_count} that the observations have caught
ULX5 at least two distinct flux levels --- a low flux level in 2001, 2014, and 2018 Feb., and a high 
flux level in 2013 and 2018 Sept. The transition between the observed
flux levels is accompanied by a change in the 0.5--10~keV
short timescale variability amplitude, which is larger than 10\% at low flux, 
but drops to 2\% at high flux. We additionally claim here that \emph{XMM-16}
represents an intermediate flux state; we justify this conclusion based on broadband spectral analysis of all data sets
in $\S$5, below. 
	
\section{Time averaged spectral analysis} 
In the next step we performed time-averaged spectral analysis. Other than the 
\emph{Suzaku-06} data, which are divided into two time intervals
(see $\S$\ref{obs:suz}), all spectra are averaged over the whole exposure. 
Each spectrum was grouped to a minimum of 20 counts in each energy bin. 
For the spectral modeling we used 
the {\sc xspec} v12.10.1 \citep{arnaud1996}, and for calculating correlations 
between parameters we used {\sc python} environment. 
Spectral fitting was performed simultaneously for all 
detectors, using constant terms for cross calibration uncertainties, 
and keeping them free 
for different detectors but fixed to unity for EPIC-pn. 
For observations taken in 2013 and 2016, the \emph{XMM-Newton} and \emph{NuSTAR} data are simultaneous, thus allowing us to put 
better constraints on the continuum model between 0.5 and 20 keV.
To account for Galactic absorption, all models 
are convolved with \texttt{tbabs} model available in {\sc xspec} with updated solar abundances 
\citep{wilms2000}. The fitting was performed using the $\chi^2$ statistic. The errors for each 
parameter are given at the 90\% confidence level, allowing the other parameters to vary during error 
calculations. Whenever we fit several model components, 
we provide the normalization of each model component (see 
Tabs.~\ref{tab:spec} and \ref{tab:acccol}). 
Note that for each model component the normalization is defined 
in different ways and carries different physical meaning and units (see the {\sc xspec} 
manual\footnote{https://heasarc.gsfc.nasa.gov/xanadu/xspec/manual/}).
 
\subsection{Spectral state transitions} 

\begin{figure}[]
    	\centering
    	\includegraphics[trim={0 0 0 0},width=0.48\textwidth]{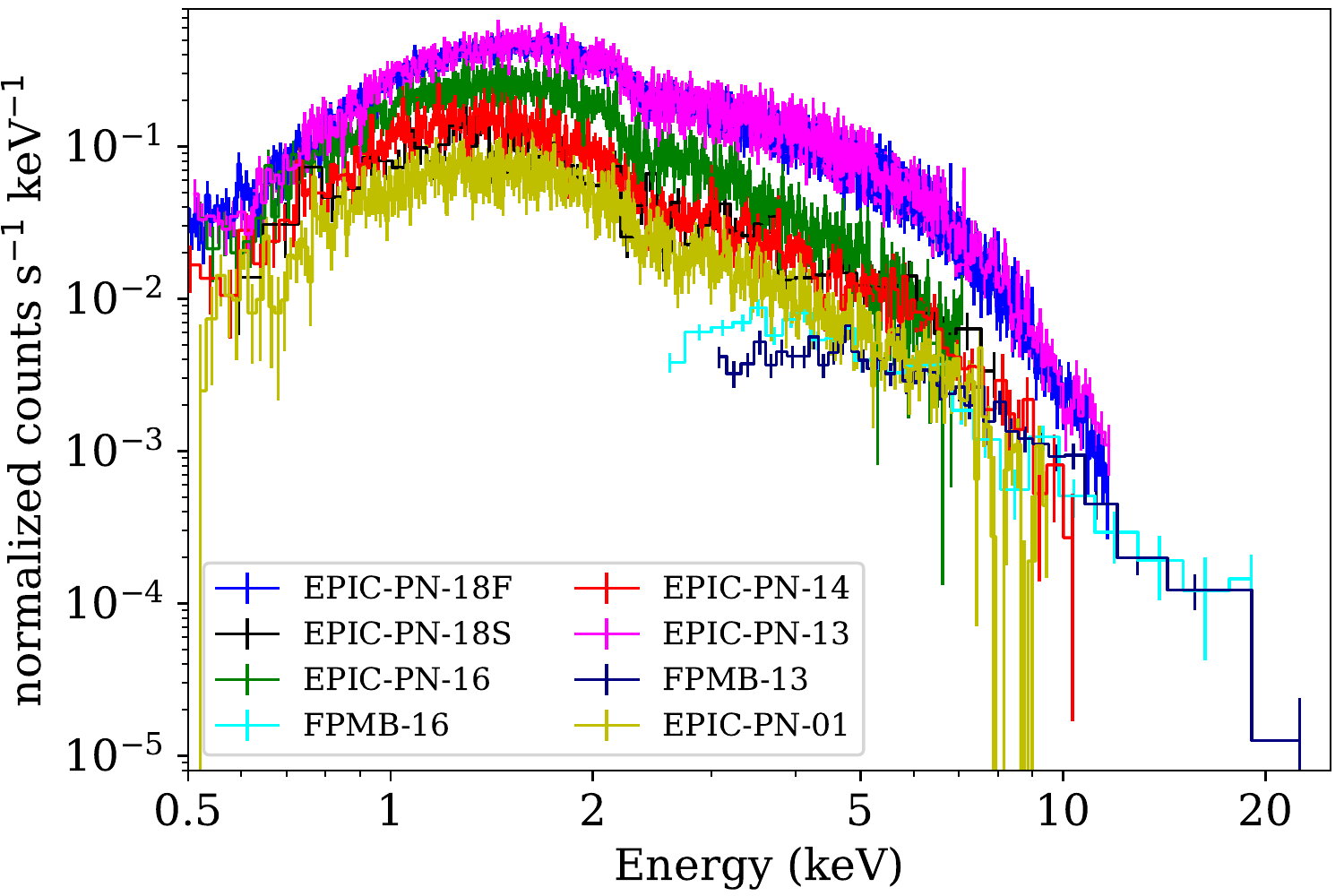}
    	\caption{Counts spectra from different epochs of observations obtained with the \emph{XMM-Newton} 
    	EPIC-pn and \emph{NuSTAR} FPMB detectors. The same spectral evolution is observed as in Fig.~\ref{fig:hr_count}.}
    	\label{fig:count_spec}
	\end{figure}

Fig.~\ref{fig:count_spec} shows the \emph{XMM-Newton} EPIC-pn and
\emph{NuSTAR} FPMB count spectra for different observational epochs
for the purpose of comparing pn to pn and FPMB to FPMB only.  We did
not include here the {\it Suzaku-06} data set, since the difference
between energy-dependent effective areas of the instruments does not
allow us to trace spectral evolution in units of normalized counts.
The color-code is the same as in Fig.~\ref{fig:hr_count}.  Based on
\emph{XMM-Newton} data in Fig.~\ref{fig:count_spec}, the same
evolution is observed as in the hardness ratio intensity diagram,
i.e., a low flux level in 2001, 2014, and 2018 Feb., an intermediate
flux level in 2016, and a high flux level in 2013 and 2018 Sept.
	
	\begin{figure}[]
    	\includegraphics[trim={0 0 0 0},width=0.48\textwidth]{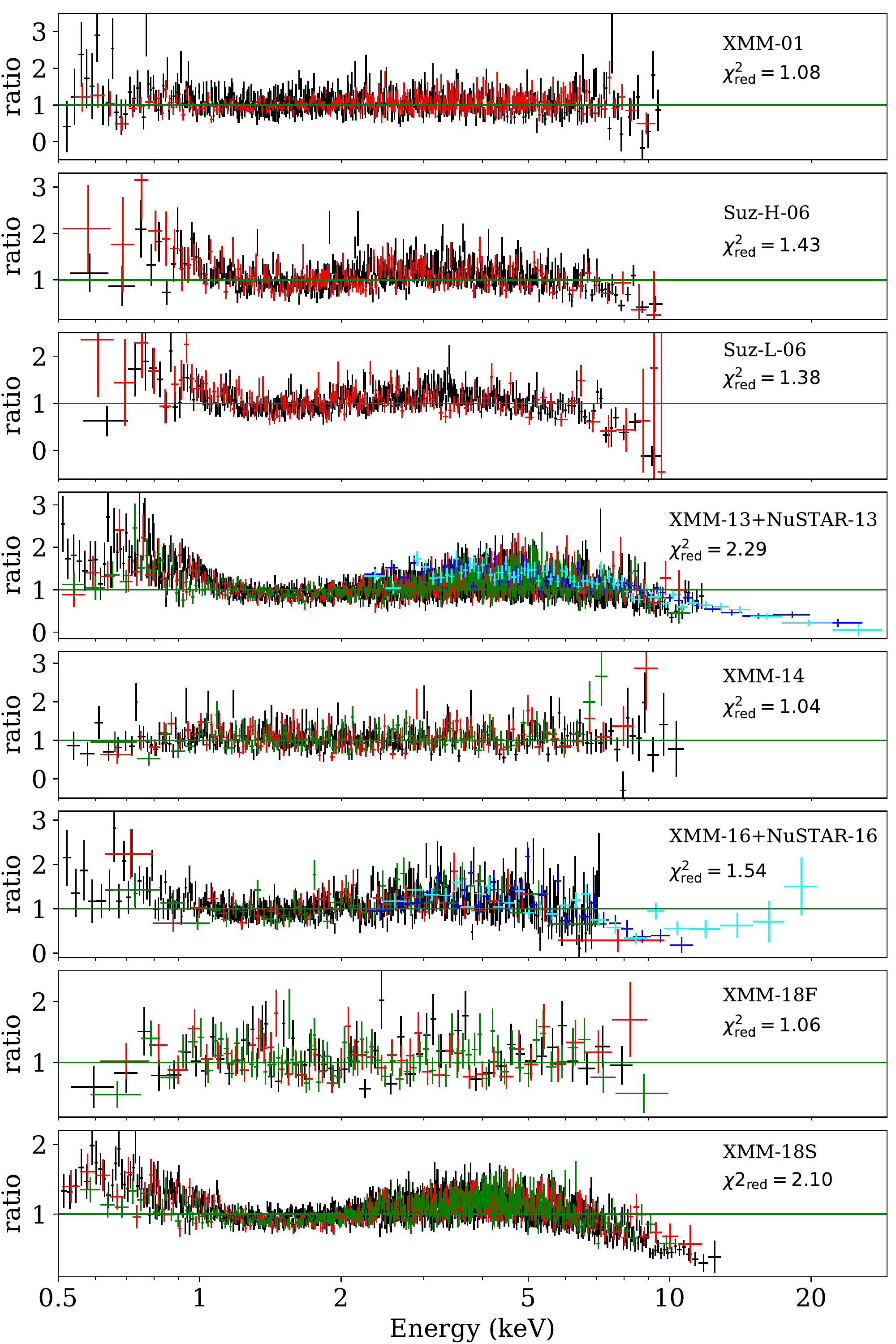}
    	\caption{The data to model ratio obtained from the fit of the \texttt{tbabs*pl} model to the 
    	data from all epochs marked at each panel. In the case of {\it XMM-Newton}, black, red and green 
    	data points are from the EPIC-pn, MOS1 and MOS2 detectors, respectively. In the case of {\it Suzaku} 
    	data, black points are from the front-illuminated XIS023 chips, and red points are from the 
    	back-illuminated XIS1 chip. The blue and cyan points are from {\it NuSTAR} FPMA and FPMB, 
    	respectively.}
    	\label{fig:xmm1_ratio}
	\end{figure}
	
	\begin{figure}[]
    	\includegraphics[trim={0 0 0 0},width=0.48\textwidth]{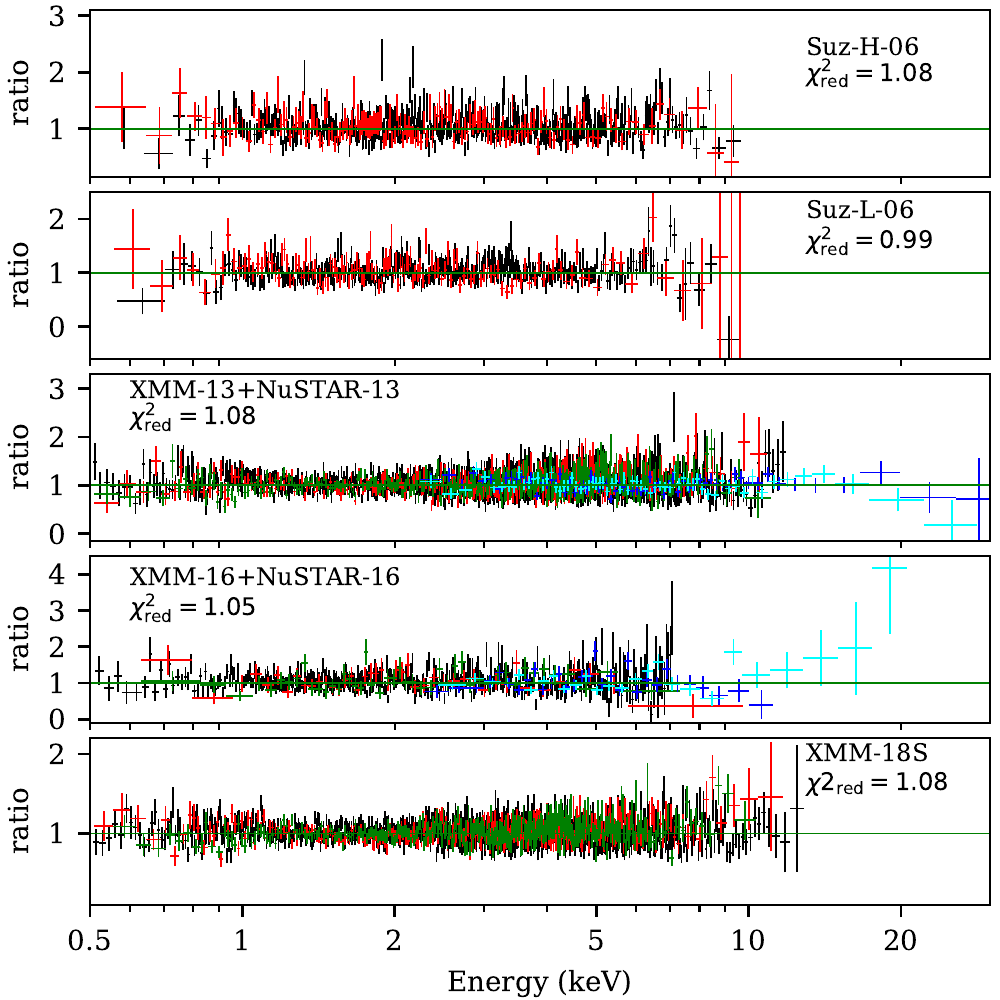}
    	\caption{The data to model ratio obtained from the \texttt{tbabs*(diskbb+pl)} model fit 
    	for the high and intermediate flux levels. A significant improvement in the residuals is visible 
    	compared to a single power-law model fit.}
    	\label{fig:bbpl_ratio}
	\end{figure}  
	
A simple absorbed power-law model (hereafter \texttt{pl}) reveals
curvature in the 2--10 keV band for data sets \emph{Suz-H-06},
\emph{Suz-L-06},\emph{XMM-13+NuSTAR+13}, \emph{XMM-16+NuSTAR+16},and
\emph{XMM-18S}, as shown in Fig.~\ref{fig:xmm1_ratio}, where the data
to model ratios for all observations are shown. This kind of
``M-shaped'' curvature has been seen in many other luminous ULXs
\citep{stobbart2006,gladstone2009}. The curvature is present even if
we allowed for a larger column density, significantly in excess of the
Galactic value, which indicates the spectra are intrinsically curved.
The curved residuals in the soft band clearly visible when ULX5 is at
the intermediate and high flux levels (Fig.~\ref{fig:xmm1_ratio}) can
be successfully modeled by a single disk component.  Moreover, data
sets at a low flux level, \emph{XMM-01}, \emph{XMM-14}, and
\emph{XMM-18F}, are well fitted by the \texttt{pl} model and do not
indicate more complex curvature.
	
To model the spectral curvature over epochs, we fit a model comprised
of two components, i.e.\ the sum of a MCD model component represented
by \texttt{diskbb} \citep{mitsuda1984,makishima1986} and the
\texttt{pl} model.  All model parameters obtained from this fit are
given in Tab.~\ref{tab:spec}, where the fit statistic is given in the
last column. We can observe a substantial fit improvement in cases
where ``M-shaped'' data are presented in Fig.~\ref{fig:xmm1_ratio},
namely for \emph{Suz-H-06}, \emph{Suz-L-06}, \emph{XMM-13+NuSTAR-13},
and \emph{XMM-18S}; values of $\chi^2_{\rm red}$ have dropped from
$\geq1.38$ to $\leq1.08$.  The improved residuals are displayed in
Fig.~\ref{fig:bbpl_ratio}.  Only \emph{XMM-16+NuSTAR-16} data still
displays a hard energy excess after removing the soft energy spectral
curvature.  Further discussion of this data set is presented in
Sec.~\ref{sec:xmm16}.

\begin{figure}[]
   		\includegraphics[trim={0 0 0 0},width=0.48\textwidth]{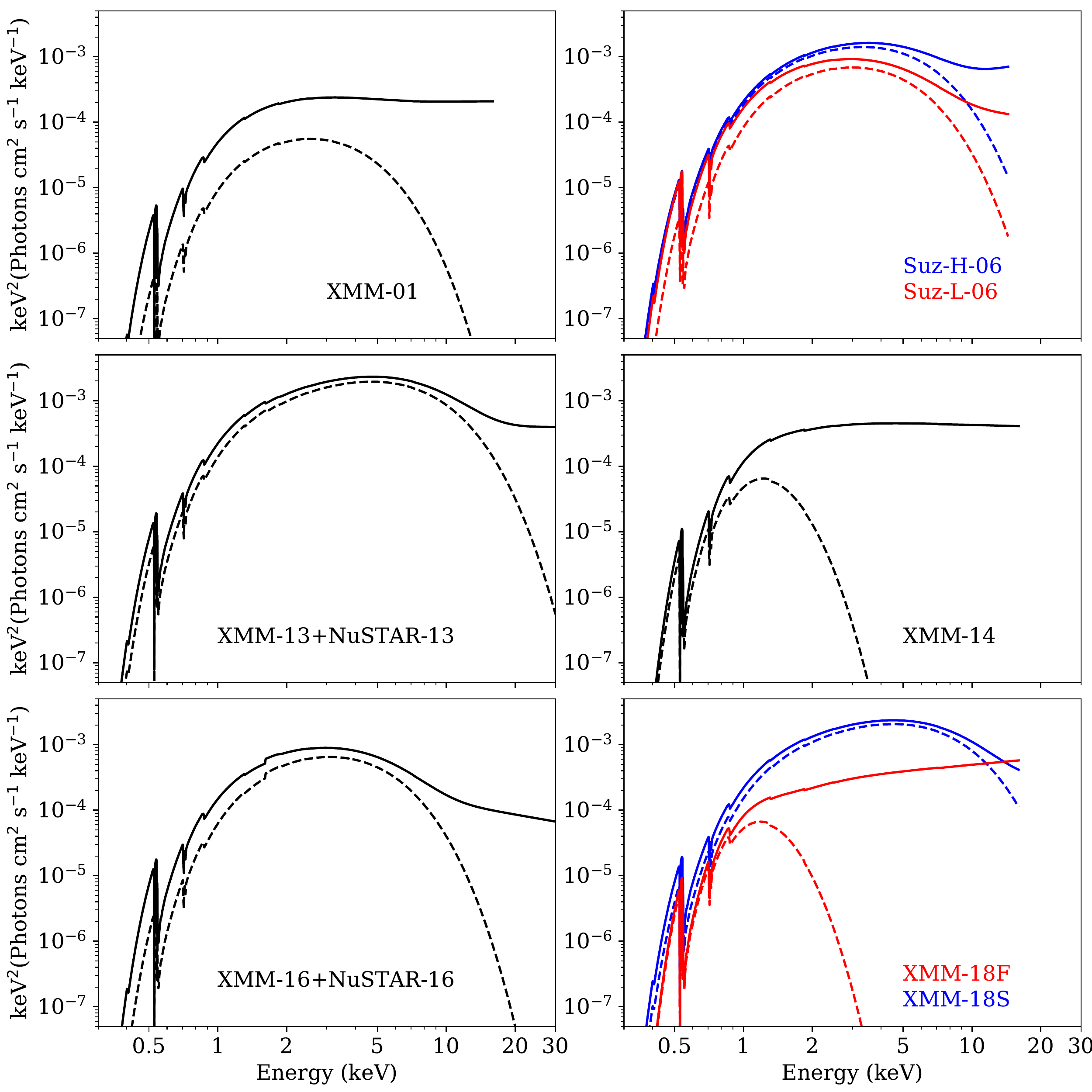}
   		\caption{The absorbed total model \texttt{diskbb+pl} (solid lines) fitted to the data from each epoch for the classification 
   		of spectral states. The disk component, \texttt{diskbb}, is also plotted with a dashed line. The source clearly evolves 
   		between a \texttt{pl} dominated state observed during 2001, 2014 and 2018 Feb.\ and a disk-dominated thermal state 
   		observed in 2006, 2013, 2016 and 2018 Sept.}
   		\label{model:state}
	\end{figure}

To display the long-term spectral evolution of Circinus ULX5,  we plot the 
absorbed best-fit model for each epoch in Fig.~\ref{model:state}. It is clearly visible that the
\emph{XMM-01}, \emph{XMM-14} and 
\emph{XMM-18F} data sets appear to be flat, consistent with an absorbed \texttt{pl}-like continuum, with 
negligible contribution from the disk. Those data sets are located in the lower island of the HID (Fig.~\ref{fig:hr_count}) 
and indicate rather high fractional variability amplitudes, $F_{\rm var}  > 10\%$ (Tab.~\ref{tab:var}).  The total 
unabsorbed X-ray luminosity for those observations ranges from 4.45 up to $7.3 \times 10^{39} $\,erg\,s$^{-1}$, 
placing these observations in a hard/low UL state.

In contrast, other spectra such as those for the \emph{Suz-H-06}, \emph{Suz-L-06}, \emph{XMM-13+NuSTAR-13}, \emph{XMM-16+NuSTAR-16},
and \emph{XMM-18S} data sets clearly show disk-dominated spectra with
higher luminosities. 
But again we can divide them onto two groups: one with luminosities ranging from 7.7 up to 
$11 \times 10^{39} $\,erg\,s$^{-1}$ in the cases of \emph{XMM-16+NuSTAR-16}, \emph{Suz-L-06} and \emph{Suz-H-06}, 
and from 16 up to $18.0  \times 10^{39} $\,erg\,s$^{-1}$ in the cases of \emph{XMM-13+NuSTAR-13} and \emph{XMM-18S}. 
We suggest that the first group represents the so-called intermediate soft state between high and low states, and the 
second group represents a typical high/soft UL state. 
	
	\begin{figure}[]
   		\includegraphics[trim={0 0 0 0},width=0.48\textwidth]{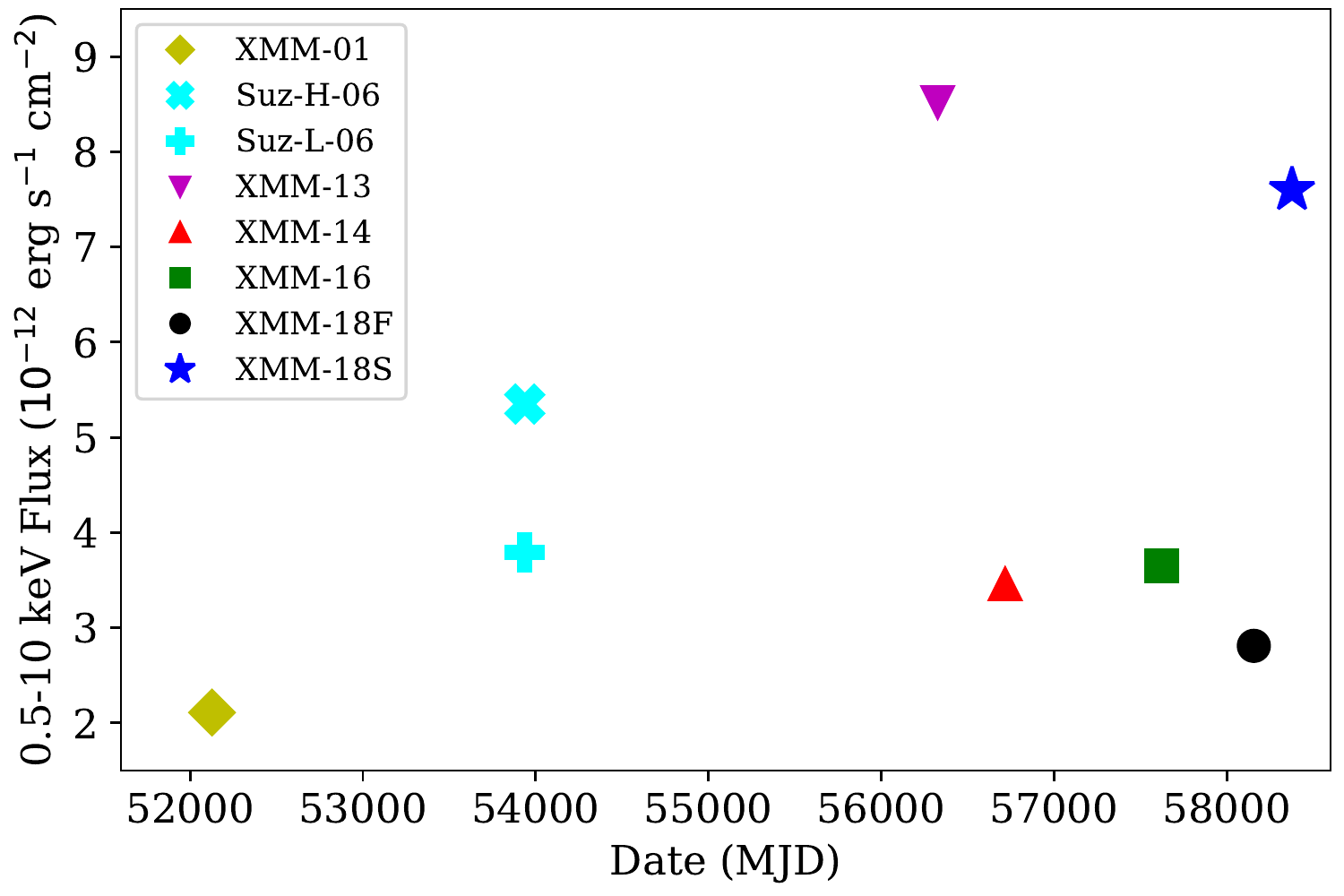}
   		\caption{Multi-epoch long-term flux variability  of Circinus ULX5, for all data analyzed in this paper.}
   		\label{fig:varflux}
	\end{figure}
	
The evolution of the unabsorbed flux  of Circinus ULX5, from the fitting of \texttt{diskbb+pl} model, is presented 
in Fig.~\ref{fig:varflux}. The evolution 
goes from the high state in 2013 to the low state in 2014, then to the
intermediate state in 2016, to the low state in 2018 Feb., and to the high state in 2018 Sept.,
with all the observed states occurring within $\sim$5 years.
The smallest observed timescale for a spectral state transition, is seven months.
Another relatively fast transition is when the source transitions from high (2013) to low (2014) 
within 11 months. However, additional multi-epoch data obtained with a more dense sampling are needed to put better 
constraints on rapid transition timescales. 
		
	\begin{table*}[]
	\tiny
    \centering
    \caption{Best fit parameters obtained from the fits of each set of data. 
    The Galactic absorption hydrogen column density is given in the second column. 
   The parameter $p$ is the index of the disk temperature radius relation, 
   $T(r) \propto r^{-p}$. $\Gamma$ is the power-law photon index. 
     For the \texttt{nthcomp} model, we tie the disk inner radius temperature 
     to the temperature of the soft seed photons which are scattered in the thermal 
    Compton process. $kT_{\rm e}$ is electron temperature of the 
    Compton-heated region. In the case of \texttt{diskbb} and \texttt{diskpbb}, the normalization is unitless:
    $(r_{\rm in}/D_{10})^2\cos(\theta)$. The normalization of the 
    \texttt{pl} is ph~keV$^{-1}$~cm$^{-2}$~s$^{-1}$  at 1 keV. The flux given in ninth column is unabsorbed.}
    \setlength{\tabcolsep}{1.9pt}               	
    \renewcommand{\arraystretch}{1.7}			
    \begin{tabular}{lrrcccccccc}
        \hline\hline
        Data & $N_{\rm H}$ & $kT_{\rm in}^{\rm bb}$ & \multirow{2}{*}{$p$} & \multirow{2}{*}{$\Gamma$} & $kT_{\rm in}^{\rm pbb}/kT_{\rm e}$ & \multirow{2}{*}{$N_{\rm norm1}$} & 
        \multirow{2}{*}{$N_{\rm norm2}$}
        & $F_{(0.5-10){\rm keV}}$ & $L_{(0.5-10){\rm keV}}$ & \multirow{2}{*}{$\chi^2/\rm DOF(\chi^2_{\rm red})$} \\
        \texttt{tbabs}*Model & $10^{21}$~[cm$^{-2}$] & [keV] &&& [keV]  &&& [erg s$^{-1}$ cm$^{-2}$]& [erg s$^{-1}$] & \\
        \hline
        
        \emph{XMM-01} & &&&&\\
        \hline
        \texttt{diskbb+pl} & $6.39^{+0.67}_{-0.71}$ & $0.90^{+0.13}_{-0.14}$ && $1.99^{+0.21}_{-0.26}$ && $1.59^{+2.46}_{-1.39}\times10^{-2}$ & $2.01^{+0.73}_{-0.77}\times10^{-4}$ & 
        $2.11\times 10^{-12}$ & $4.45\times 10^{39}$ & 748/703(1.06) \\
        \texttt{diskpbb} & $6.80^{+0.15}_{-0.28}$ && $0.50^{+0.01}_{-0.01}$ && $2.77^{+0.27}_{-0.41}$ & $1.69^{+1.11}_{-0.74}\times10^{-4}$ && $2.16\times 10^{-12}$ & $4.56\times 10^{39}$ & 
        752/704(1.07) \\
		\texttt{diskbb+nthcomp} & $5.19^{+0.26}_{-0.23}$ & $0.77^{+0.07}_{-0.01}$ && $1.28^{+1.15}_{-0.01}$ & $1.62^{+6.07}_{-0.17}$ & $9.60^{+17.20}_{-2.73}\times10^{-2}$ & 
		$3.63^{+0.15}_{-0.15}\times10^{-6}$ & $1.80\times 10^{-12}$ & $3.80\times 10^{39}$ & 743/702(1.06) \\
		\texttt{diskbb+diskpbb} & $5.57^{+1.11}_{-0.34}$ & $0.66^{+0.09}_{-0.10}$ & $0.61^{+0.05}_{-0.05}$ && $2.79^{+0.28}_{-0.19}$ & $1.11^{+0.40}_{-0.76}\times10^{-1}$ & $3.29^{+1.60}_{-1.14}\times10^{-4}$ & 
		$1.86\times10^{-12}$ & $3.92\times 10^{39}$ & 744/702(1.06) \\
        \hline
        
        
        \emph{Suz-H-06} & &&&&\\
        \hline
        \texttt{diskbb+pl} & $5.32^{+0.94}_{-0.42}$ & $1.29^{+0.08}_{-0.06}$ && $1.23^{+0.14}_{-0.15}$ && $8.02^{+1.63}_{-2.32}\times10^{-2}$ & $8.78^{+1.92}_{-1.76}\times10^{-5}$ & 
        $5.35\times 10^{-12}$ & $1.13\times 10^{40}$ & 599/553(1.08) \\
        \texttt{diskpbb} & $6.04^{+0.69}_{-0.68}$ && $0.63^{+0.05}_{-0.04}$ && $1.61^{+0.11}_{-0.10}$ & $2.50^{+1.41}_{-0.91}\times10^{-2}$ && $5.62\times 10^{-12}$ & $1.19\times 10^{40}$ & 
        607/554(1.10) \\
		\texttt{diskbb+nthcomp} & $4.28^{+0.78}_{-0.99}$ & $0.61^{+0.23}_{-0.12}$ && $2.41^{+0.17}_{-0.15}$ & $2.26^{+0.24}_{-0.19}$ & $2.76^{+0.87}_{-1.21}\times10^{-1}$ & 
		$3.88^{+0.35}_{-0.34}\times10^{-4}$ & $4.97\times 10^{-12}$ & $1.05\times 10^{40}$ & 594/552(1.08) \\
		\texttt{diskbb+diskpbb} & $7.07^{+0.55}_{-0.52}$ & $0.14^{+0.07}_{-0.05}$ & $0.61^{+0.03}_{-0.02}$ && $1.65^{+0.10}_{-0.09}$ & $331.86^{+189.03}_{-190.08}$ & $2.03^{+0.79}_{-0.58}\times10^{-2}$ & 
		$6.52\times10^{-12}$ & $1.38\times 10^{40}$ & 602/552(1.09) \\
        \hline
        
        \emph{Suz-L-06} & &&&&\\
        \hline
        \texttt{diskbb+pl} & $6.45^{+0.13}_{-0.13}$ & $1.11^{+0.09}_{-0.08}$ && $2.43^{+0.18}_{-0.17}$ && $7.61^{+3.43}_{-2.23}\times10^{-2}$ & $4.17^{+3.49}_{-3.11}\times10^{-4}$ & 
        $3.79\times 10^{-12}$ & $7.99\times 10^{39}$ & 458/461(0.99) \\
        \texttt{diskpbb} & $6.38^{+0.68}_{-0.68}$ && $0.57^{+0.04}_{-0.03}$ && $1.35^{+0.09}_{-0.08}$ & $2.22^{+1.31}_{-0.83}\times10^{-2}$ && $3.66\times 10^{-12}$ & $7.72\times 10^{39}$ & 
        459/462(0.99) \\
		\texttt{diskbb+nthcomp} & $5.12^{+0.16}_{-0.15}$ & $0.72^{+0.16}_{-0.28}$ && $2.28^{+0.15}_{-0.06}$ & $1.62^{+0.36}_{-0.20}$ & $3.42^{+0.42}_{-0.10}\times10^{-1}$ & 
		$1.08^{+0.20}_{-0.01}\times10^{-4}$ & $3.21\times 10^{-12}$ & $6.77\times 10^{39}$ & 454/460(0.98) \\
		\texttt{diskbb+diskpbb} & $7.56^{+0.52}_{-0.49}$ & $0.15^{+0.06}_{-0.04}$ & $0.55^{+0.03}_{-0.02}$ && $1.38^{+0.09}_{-0.08}$ & $259.69^{+134.99}_{-136.57}$ & $1.82^{+0.80}_{-0.57}\times10^{-2}$ & 
		$4.56\times10^{-12}$ & $9.62\times 10^{39}$ & 453/460(0.98) \\
        \hline
        
        \emph{XMM-13+NuSTAR-13} & &&&&\\
        \hline
        \texttt{diskbb+pl} & $6.40^{+0.33}_{-0.33}$ & $1.88^{+0.05}_{-0.04}$ && $2.01^{+0.12}_{-0.12}$ && $2.43^{+0.27}_{-0.24}\times10^{-2}$ & $4.70^{+1.15}_{-1.12}\times10^{-4}$ & 
        $8.52\times 10^{-12}$ & $1.80\times 10^{40}$ & 2009/1859(1.08) \\
        \texttt{diskpbb} & $6.29^{+0.21}_{-0.21}$ && $0.64^{+0.01}_{-0.01}$ && $2.24^{+0.06}_{-0.06}$ & $9.88^{+1.61}_{-1.29}\times10^{-3}$ && $8.44\times 10^{-12}$ & $1.78\times 10^{40}$ & 
        2069/1860(1.11) \\
		\texttt{diskbb+nthcomp} & $5.33^{+0.12}_{-0.12}$ & $1.68^{+0.06}_{-0.24}$ && $\sim1.00$ & $2.95^{+1.41}_{-0.33}$ & $4.44^{+2.23}_{-0.47}\times10^{-2}$ & $1.33^{+0.14}_{-0.15}\times10^{-6}$ & 
		$7.88\times 10^{-12}$ & $1.66\times 10^{40}$ & 2018/1858(1.08) \\
		\texttt{diskbb+diskpbb} & $7.04^{+0.18}_{-0.17}$ & $0.18^{+0.04}_{-0.03}$ & $0.62^{+0.01}_{-0.01}$ && $2.26^{+0.06}_{-0.06}$ & $56.94^{+14.56}_{-14.63}$ & $9.10^{+1.52}_{-1.24}\times10^{-3}$ & 
		$9.08\times10^{-12}$ & $1.92\times 10^{40}$ & 2045/1858(1.10) \\
        \hline
        
        \emph{XMM-14}  & &&&&\\
        \hline
        \texttt{diskbb+pl} & $8.25^{+1.62}_{-1.32}$ & $0.20^{+0.08}_{-0.04}$ && $2.11^{+0.11}_{-0.11}$ && $68.82^{+33.90}_{-31.72}$ & $5.55^{+1.00}_{-0.87}\times10^{-4}$ & 
        $3.46\times 10^{-12}$ & $7.30\times 10^{39}$ & 477/467(1.02) \\
        \texttt{diskpbb} & $5.93^{+0.17}_{-0.18}$ && $0.50^{+0.02}_{-0.02}$ && $3.55^{+0.62}_{-0.49}$ & $1.16^{+0.56}_{-0.50}\times10^{-4}$ && $2.36\times 10^{-12}$ & $4.98\times 10^{39}$ & 
        498/468(1.06) \\
		\texttt{diskbb+nthcomp} & $7.85^{+1.67}_{-1.51}$ & $0.21^{+0.07}_{-0.04}$ && $2.06^{+0.09}_{-0.09}$ & $5.61^{+2.43}_{-2.43}$ & $63.93^{+28.34}_{-27.74}$ & 
		$4.32^{+1.22}_{-1.31}\times10^{-4}$ & $3.15\times 10^{-12}$ & $6.65\times 10^{39}$ & 477/466(1.02) \\
		\texttt{diskbb+diskpbb} & $7.87^{+1.01}_{-1.32}$ & $0.22^{+0.01}_{-0.01}$ & $0.50^{+0.01}_{-0.01}$ && $3.81^{+0.19}_{-0.05}$ & $36.30^{+15.64}_{-12.70}$ & $9.03^{+1.38}_{-0.41}\times10^{-5}$ & 
		$3.16\times10^{-12}$ & $6.67\times 10^{39}$ & 478/466(1.03) \\
        \hline
        
        \emph{XMM-16+NuSTAR-16} & &&&&\\
        \hline
		\texttt{diskbb+pl} & $6.59^{+0.81}_{-0.94}$ & $1.15^{+0.08}_{-0.07}$ && $2.61^{+0.30}_{-0.35}$ && $5.18^{+1.90}_{-1.32}\times10^{-2}$ & $4.55^{+2.11}_{-2.20}\times10^{-4}$ & 
        $3.65\times 10^{-12}$ & $7.70\times 10^{39}$ & 591/561(1.05) \\
        \texttt{diskpbb} & $6.25^{+0.48}_{-0.48}$ && $0.55^{+0.03}_{-0.02}$ && $1.44^{+0.09}_{-0.08}$ & $1.25^{+0.63}_{-0.42}\times10^{-2}$ && $3.41\times 10^{-12}$ & $7.20\times 10^{39}$ & 
        601/562(1.07) \\
		\texttt{diskbb+nthcomp} & $4.72^{+0.26}_{-0.17}$ & $1.02^{+0.15}_{-0.22}$ && $2.18^{+0.32}_{-0.25}$ & $\sim363$ & $1.07^{+2.88}_{-0.29}\times10^{-1}$ & 
		$1.22^{+8.22}_{-1.14}\times10^{-5}$ & $2.86\times10^{-12}$ & $6.03\times 10^{39}$ & 603/560(1.08) \\
		\hline

        \emph{XMM-18F} & &&&&\\
        \hline
		\texttt{diskbb+pl} & $8.47^{+2.52}_{-2.39}$ & $0.19^{+0.07}_{-0.04}$ && $1.69^{+0.15}_{-0.18}$ && $144.49^{+17.44}_{-17.04}$ & $2.47^{+0.73}_{-0.58}\times10^{-4}$ & 
        $2.81\times 10^{-12}$ & $5.93\times 10^{39}$ & 174/174(1.00) \\
        \texttt{diskpbb} & $4.18^{+0.53}_{-0.51}$ && $0.56^{+0.02}_{-0.01}$ && $4.15^{+0.12}_{-0.12}$ & $8.96^{+1.92}_{-1.64}\times10^{-5}$ && $1.54\times 10^{-12}$ & $3.25\times 10^{39}$ & 
        186/175(1.06) \\
		\texttt{diskbb+nthcomp} & $8.15^{+2.10}_{-2.23}$ & $0.19^{+0.06}_{-0.04}$ && $1.68^{+0.13}_{-0.17}$ & $\sim994$ & $113.20^{+13.42}_{-13.12}$ & 
		$2.19^{+0.86}_{-0.46}\times10^{-4}$ & $2.62\times10^{-12}$ & $5.53\times 10^{39}$ & 174/173(1.01) \\
		\texttt{diskbb+diskpbb} & $7.13^{+0.45}_{-0.41}$ & $0.23^{+0.01}_{-0.02}$ & $0.60^{+0.05}_{-0.04}$ && $3.35^{+0.27}_{-0.23}$ & $24.34^{+12.43}_{-10.63}$ & $2.84^{+1.31}_{-0.80}\times10^{-4}$ & 
		$2.10\times10^{-12}$ & $4.43\times 10^{39}$ & 174/173(1.01) \\
		\hline
        
        \emph{XMM-18S} & &&&&\\
        \hline
		\texttt{diskbb+pl} & $5.95^{+0.62}_{-0.37}$ & $1.83^{+0.06}_{-0.07}$ && $2.00^{+0.59}_{-0.34}$ && $2.83^{+0.41}_{-0.27}\times10^{-2}$ & $3.05^{+1.57}_{-1.12}\times10^{-4}$ & 
        $7.60\times 10^{-12}$ & $1.60\times 10^{40}$ & 2324/2146(1.08) \\
        \texttt{diskpbb} & $5.68^{+0.15}_{-0.15}$ && $0.68^{+0.01}_{-0.01}$ && $1.98^{+0.04}_{-0.04}$ & $1.89^{+0.26}_{-0.23}\times10^{-2}$ && $7.43\times10^{-12}$ & $1.57\times 10^{40}$ & 
        2330/2147(1.09) \\
		\texttt{diskbb+nthcomp} & $5.37^{+0.08}_{-0.07}$ & $1.32^{+0.23}_{-0.30}$ && $\sim 1.10$ & $1.73^{+0.45}_{-0.16}$ & $8.77^{+8.56}_{-3.33}\times10^{-2}$ & 
		$1.57^{+6.83}_{-1.10}\times10^{-5}$ & $7.28\times10^{-12}$ & $1.54\times 10^{40}$ & 2331/2145(1.09) \\
		\texttt{diskbb+diskpbb} & $7.68^{+0.27}_{-0.31}$ & $0.19^{+0.01}_{-0.01}$ & $0.65^{+0.01}_{-0.01}$ && $1.99^{+0.04}_{-0.04}$ & $129.17^{+52.14}_{-53.97}$ & $1.61^{+0.20}_{-0.18}\times10^{-2}$ & 
		$8.95\times10^{-12}$ & $1.89\times 10^{40}$ & 2320/2145(1.08) \\
		\hline
        
    \end{tabular}
    \label{tab:spec}
	\end{table*}

\subsection{Accretion mode in ULX5} 
We also check if the slim disk model,  \texttt{diskpbb} \citep{abramowicz88,watarai2000}, 
can improve our fit, thus indicating nearly-Eddington accretion.
The slim disk model gives 
a flatter effective temperature radial profile, i.e.\ the index $p$ in the
 relation $T(r) \propto r^{-p}$ is lower than 0.75, the latter being 
 a typical value for the standard accretion disk SS73 model. 
 For each epoch of our data we obtain  the best-fit value for the $p$ parameter
 as listed in Tab.~\ref{tab:spec}. These values are all $<0.75$. 
 Interestingly, when the source is in the low state (\emph{XMM-01}, 
 \emph{XMM-14}, \emph{XMM-18F}) $p$ is $0.50\pm 0.01$, $0.50\pm 0.02$ 
and $0.55^{+0.02}_{-0.01}$, respectively; for the intermediate state  
(\emph{Suz-L-06}, \emph{Suz-H-06}, \emph{XMM-16}), $p$ is 
$0.57^{+0.04}_{-0.03}$,   $0.63^{+0.05}_{-0.04}$ and $0.55^{+0.03}_{-0.02}$, respectively; for the high state 
 (\emph{XMM-13}, \emph{XMM-18S}), $p$ is $0.64 \pm 0.01$  and 
 $0.68 \pm 0.01$, respectively. 
 Such a result is consistent with the notion that 
 a super-Eddington accretion flow in an advection-dominated slim disk can 
 explain the properties of ULX5. Furthermore  we noticed that as the total flux 
 increases, the values of $p$ moves towards 0.75, the values from the standard 
 thin disk model. This movement may indicate that the slim disk model is a good description 
 for relatively low-luminosity ULXs with nearly- or above super-Eddington accretion, 
 as pointed out by \cite{sutton2013}.

Finally, we tested a two-component model accounting for thermal 
Comptonization in a hot plasma near the accretion disk, i.e.\ 
\texttt{diskbb} plus \texttt{nthcomp} \citep{zdziarski96, zycki99}. 
All model parameters obtained from this fit are given in Tab.~\ref{tab:spec}. 
In general,  two-component models such as \texttt{diskbb+pl} and \texttt{diskbb+nthcomp} 
yield slightly better test statistics than \texttt{diskpbb}, as seen at the last column of Tab.~\ref{tab:spec}. 
We checked the Akaike Information Criterion 
\citep[AIC][]{akaike1973,sugiura78} to compare the models. 
The AIC information tells us that the  \texttt{diskbb+nthcomp} model 
slightly better describes the data than  \texttt{diskbb+pl} model in case of  \emph{XMM-01},  \emph{Suz-H-06} and  \emph{Suz-L-06} data sets, 
but it is opposite in case of  \emph{XMM-13},  \emph{XMM-16} and  \emph{XMM-18S} data sets. For  \emph{XMM-14}  and \emph{XMM-18F} these two models are indistinguishable. 

Furthermore, in two cases (\emph{XMM-16+NuSTAR-16}, \emph{XMM-18F}), 
the electron temperature is not fully constrained, which may indicate that either the 
model is questionable or we have too few data points at higher energies.

To put a better constraint on the   accretion mode operating in ULX5, we determined the inner disk radius $R_{\rm in}$ 
and the corresponding inner disk temperature $T_{\rm in}$, and their evolution over different epochs. One can estimate 
the value of $R_{\rm in}$ from the normalization of the \texttt{diskbb} component, if the distance to the source and the 
viewing angle are known. The normalization of the \texttt{diskbb} component is $(r_{\rm in}/D_{10})^2\cos(\theta)$, 
where $D_{10}$ is the distance to the source in units of 10 kpc, $\theta$ is the viewing angle, and $r_{\rm in}$ is 
the apparent inner disk radius, i.e.\ without accounting for the 
color-correction factor. The inner disk radius 
after color-correction is given by $R_{\rm in}$=$\xi\kappa^2r_{\rm in}$, where we used 
$\xi$=$\sqrt{\frac{3}{7}}(\frac{6}{7})^{3}$ and $\kappa$=1.7 \citep{kubota1998}. To do so, we need the disk component to be 
strongly visible in the data, otherwise the fitting  can lead to parameter degeneracy between $T_{\rm in}$ and 	
the \texttt{diskbb} model normalization. For our data, we estimated $R_{\rm in}$ assuming the distance to the 
source is 4.2 Mpc and the source is viewed face-on (i.e. $\theta$=0).

	\begin{figure}[]
   		\centering
   		\includegraphics[trim={0 0 0 0},width=0.48\textwidth]{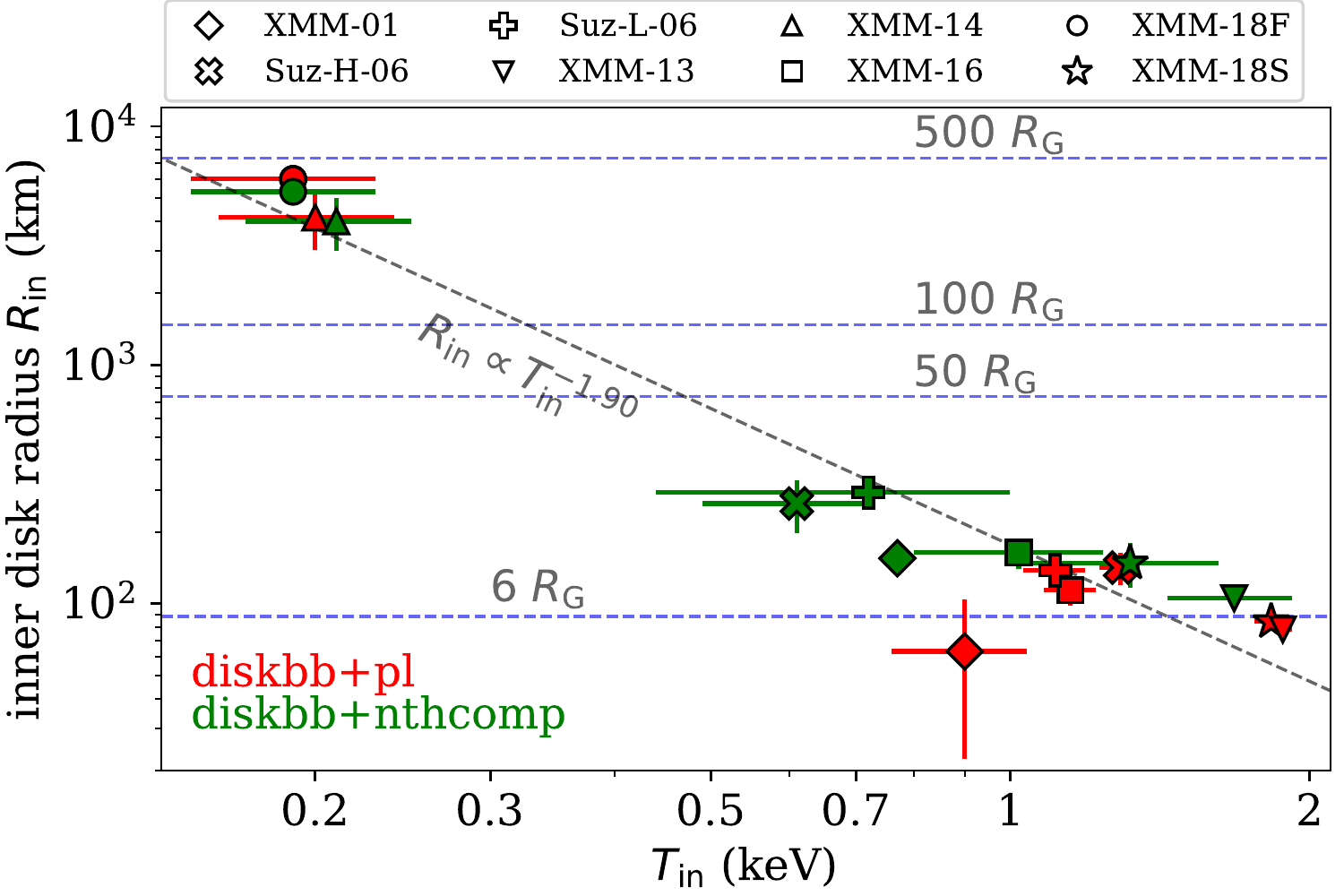}
   		\caption{The relation between disk $R_{\rm in}$ and $T_{\rm in}$ obtained from spectral fitting of two models, 
   		  \texttt{diskbb+pl} and \texttt{diskbb+nthcomp}, with data shown in red and green, respectively.
                  The relation $R_{\rm in}\propto T_{\rm in}^{-1.90\pm0.15}$ 
   		is plotted with a dashed gray line. The blue horizontal dashed lines show various radii for a 10\,M$_{\odot}$ BH. In 
   		the case of three fits, $R_{\rm in}$ is lower than the ISCO for a non-spinning BH.}
   		\label{fig:Rin_Tin}
	\end{figure}
	
	\begin{figure}[]
   		\centering
   		\includegraphics[trim={0 0 0 0},width=0.48\textwidth]{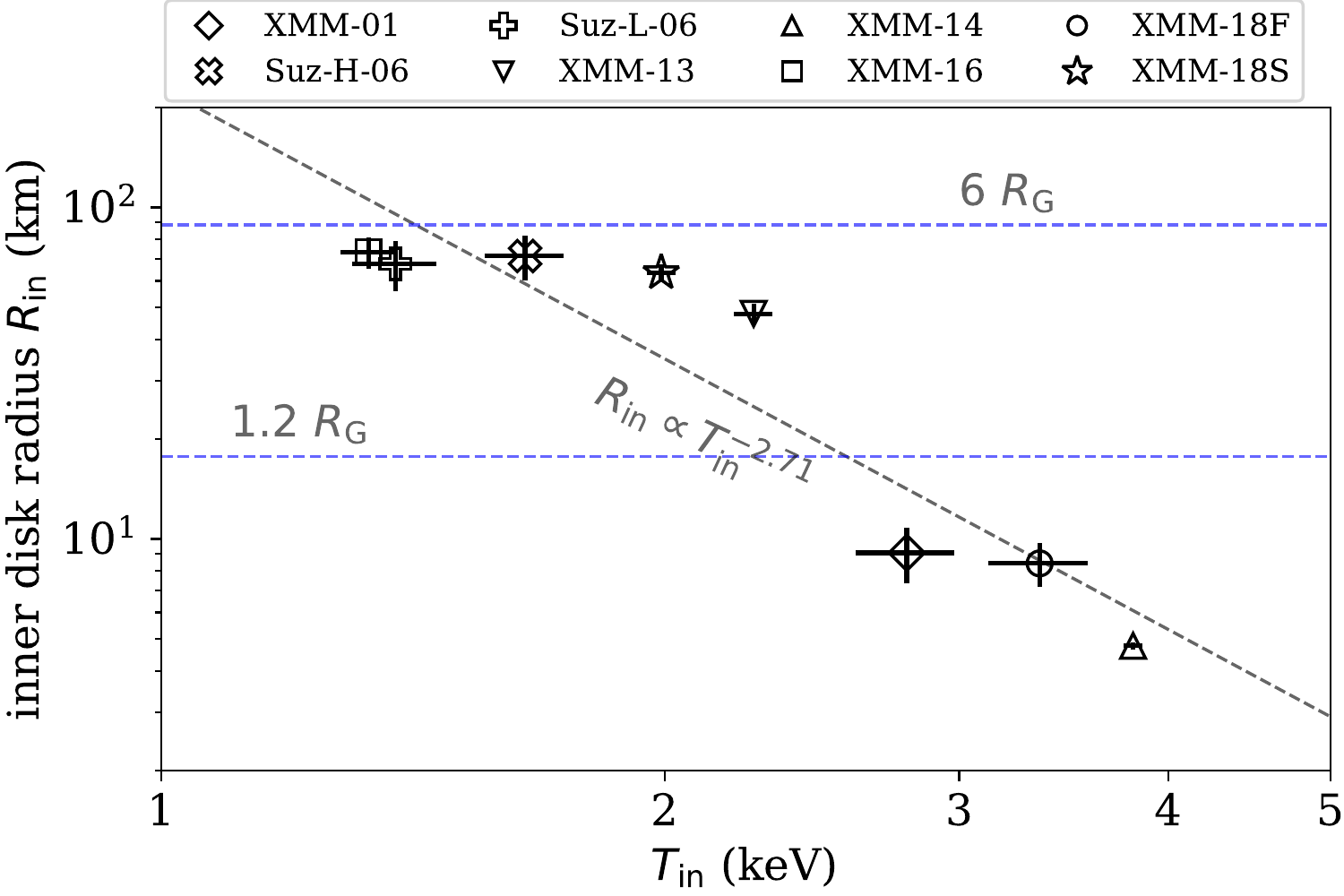}
   		\caption{The relation between disk $R_{\rm in}$ and $T_{\rm in}$
   		(i.e. $kT^{\rm pbb}_{\rm in} $ in Tab.~\ref{tab:spec}) 
   		obtained from spectral fitting of \texttt{diskbb+diskpbb}. 
   		The relation $R_{\rm in}\propto T_{\rm in}^{-2.71\pm0.45}$ is 
   		plotted with a dashed 
   		gray line. The blue horizontal dashed lines show various radii for a 10\,M$_{\odot}$ BH. In the case of three fits, $R_{\rm in}$ 
   		is lower than the ISCO for a maximally spinning BH.}
   		\label{fig:pbbRin_Tin}
	\end{figure}
	
	
	\begin{figure}[]
   		\centering
   		\includegraphics[trim={0 0 0 0},width=0.48\textwidth]{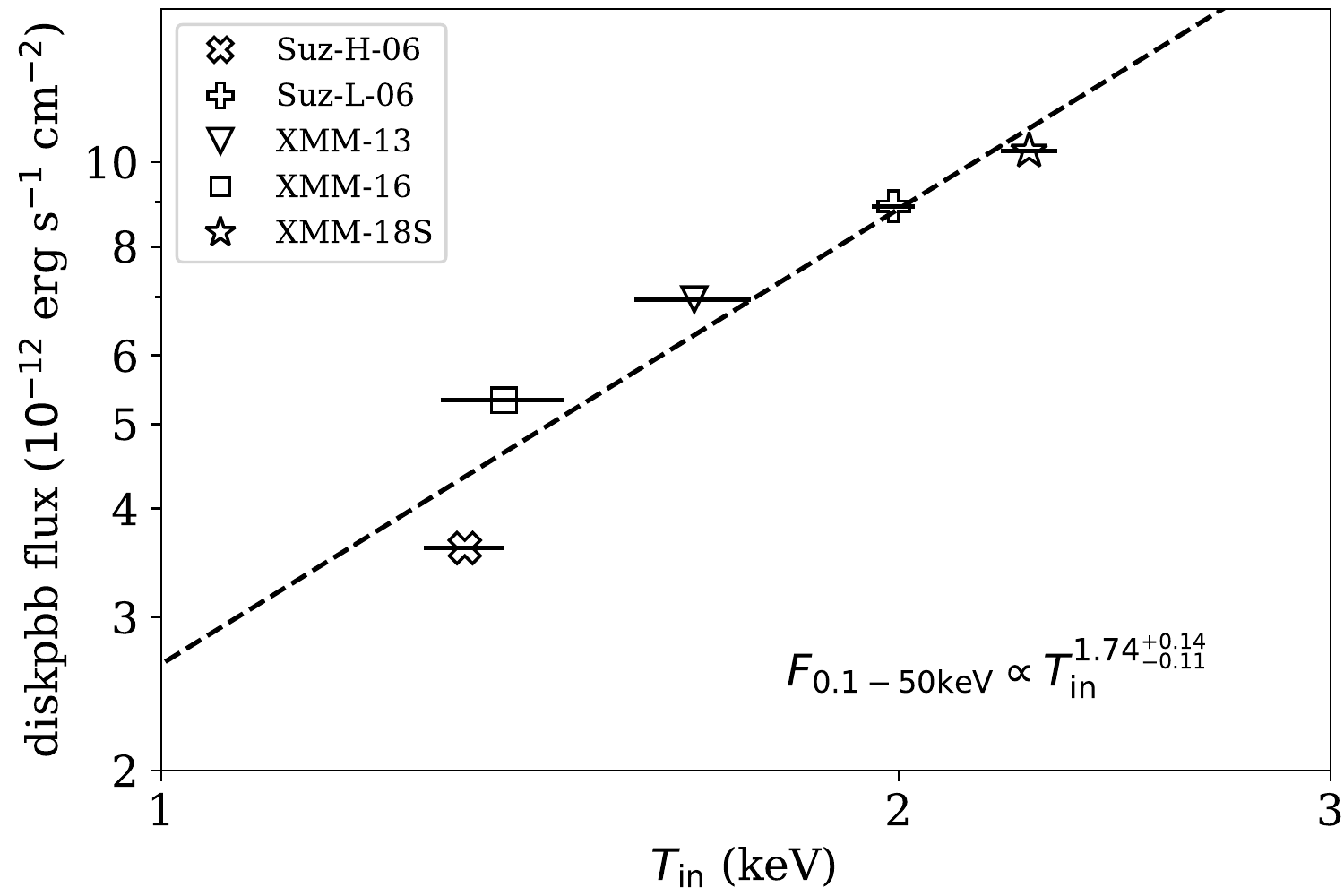}
   		\caption{The \texttt{diskpbb} flux versus $T_{\rm in}$ relation inferred from spectral fitting of \texttt{diskbb+diskpbb} model. 
   		Only soft thermal disk-dominated spectra are used. The dashed black  line shows the best fit to the data points.}
   		\label{fig:diskflux_Tin}
	\end{figure}

Fig. \ref{fig:Rin_Tin} shows the estimated inner disk radius from the \texttt{diskbb+pl} and \texttt{diskbb+nthcomp} 
model fittings. The inner disk radius and temperature follows an inverse relation, $R_{\rm in}\propto T_{\rm in}^{-1.89\pm0.15}$. 
In general, as the source flux increases the inner disk temperature increases and the inner disk radius decreases. 
We compare the inferred values of radius to the innermost stable circular orbit (ISCO) radius of a non-rotating 10\,M$_{\odot}$ 
BH. When the source is in a clearly \texttt{pl}-dominated hard/low spectral state, i.e.\ for \emph{XMM-14} and \emph{XMM-18F}, 
the inner disk radius is about 400 gravitational radii, $R_{\rm G}$, which is a very large value, but the weakness of the disk 
component in these data do not allow us to constrain this parameter well. For almost all other data sets, the disk goes 
down to the ISCO, though there are three points where,
in case of \texttt{diskbb+pl}, $R_{\rm in}$ goes below 6\,$R_{\rm G}$. 
Such a scenario can be realized in the case of a maximally pro-grade rotating BH, when ISCO radius can be brought  down from 
6 to 1.2~$R_{\rm G}$, or in the case when the mass of the central object is even lower than 10\,M$_{\odot}$. The distribution 
of inner disk radii provides the value of $T_{\rm in}$, which in disk-dominated spectral states is quite 
high, $\sim$1-2\,keV, comparable 
to values observed in XRBs, making Circinus ULX5 notable and distinct from the rest of the ULXs. 
 
Nevertheless, both models presented in Fig.~\ref{fig:Rin_Tin} represent geometrically 
thin sub-Eddington accretion disk whereas the normalization of the \texttt{diskbb} component suggests lower mass BH. 
To account for super-Eddington accretion indicated by high luminosity, we 
constructed physically more consistent model composed 
of \texttt{diskbb} component for the outer sub-Eddington part of the disk which is accounted for only the low energy curvature, plus \texttt{diskpbb} 
component for the inner super-Eddington part of the disk which fits the high energy component. The fitting parameters are given in 
Tab. \ref{tab:spec}, and in Tab.~\ref{tab:acccol} in case of 2016 data set. Then in Fig.~\ref{fig:pbbRin_Tin} we plot the inner disk radius and temperature relation from the \texttt{diskpbb} model component. 
The absence of the high energy curvature in three data sets observed during low states (\emph{XMM-01}, \emph{XMM-14} and \emph{XMM-18F}) leads to high inner disk temperature and very low inner disk radius even lower than the ISCO radius 
of a maximally spinning 10 M$_{\odot}$ BH. On the other hand, the inner disk radius obtained from the high flux data sets is lying just below 
6 $R_{\rm G}$. 

Next, in Fig.~\ref{fig:diskflux_Tin}, we plot unabsorbed 0.1--50 keV disk flux (the allowed limit in {\sc xspec} for the \texttt{diskpbb} model) 
versus inner disk temperature obtained from the same \texttt{diskpbb} model component. In the standard regime, where the value of
$R_{\rm in}$ obtained from the disk component normalization remains constant, the disk flux should follow $F_{\rm disk}\propto T_{\rm in}^4$ 
\citep{kubota2002} for thin accretion disk and $F_{\rm disk}\propto T_{\rm in}^2$ for advection dominated slim accretion disk. From 
this figure we excluded observations where $T_{\rm in}$ is not properly constrained due to the absence of a clear disk component, and which led to
the smaller values of inner disk radius shown in Fig.~\ref{fig:pbbRin_Tin}. If we fit the data points 
ignoring the \emph{XMM-01}, \emph{XMM-14} and \emph{XMM-18F} data sets, the scaling relation becomes 
$F_{\rm disk}\propto T_{\rm in}^{1.74^{+0.14}_{-0.11}}$. 
This is clearly what one would expect from the advection dominated accretion mode where the correlation power goes $\sim$2. This also demonstrates that we 
have a direct view of accreting matter not covered by the disk wind, since the flux temperature correlation power 
is positive \citep[for opposite result see:][]{feng2007,kajava2009,mondal2020b}.

\subsection{The XMM-16 and  NuSTAR-16 data set} 
\label{sec:xmm16}
\begin{figure}[]
    	\centering
    	\includegraphics[trim={0 0 0 0},width=0.48\textwidth]{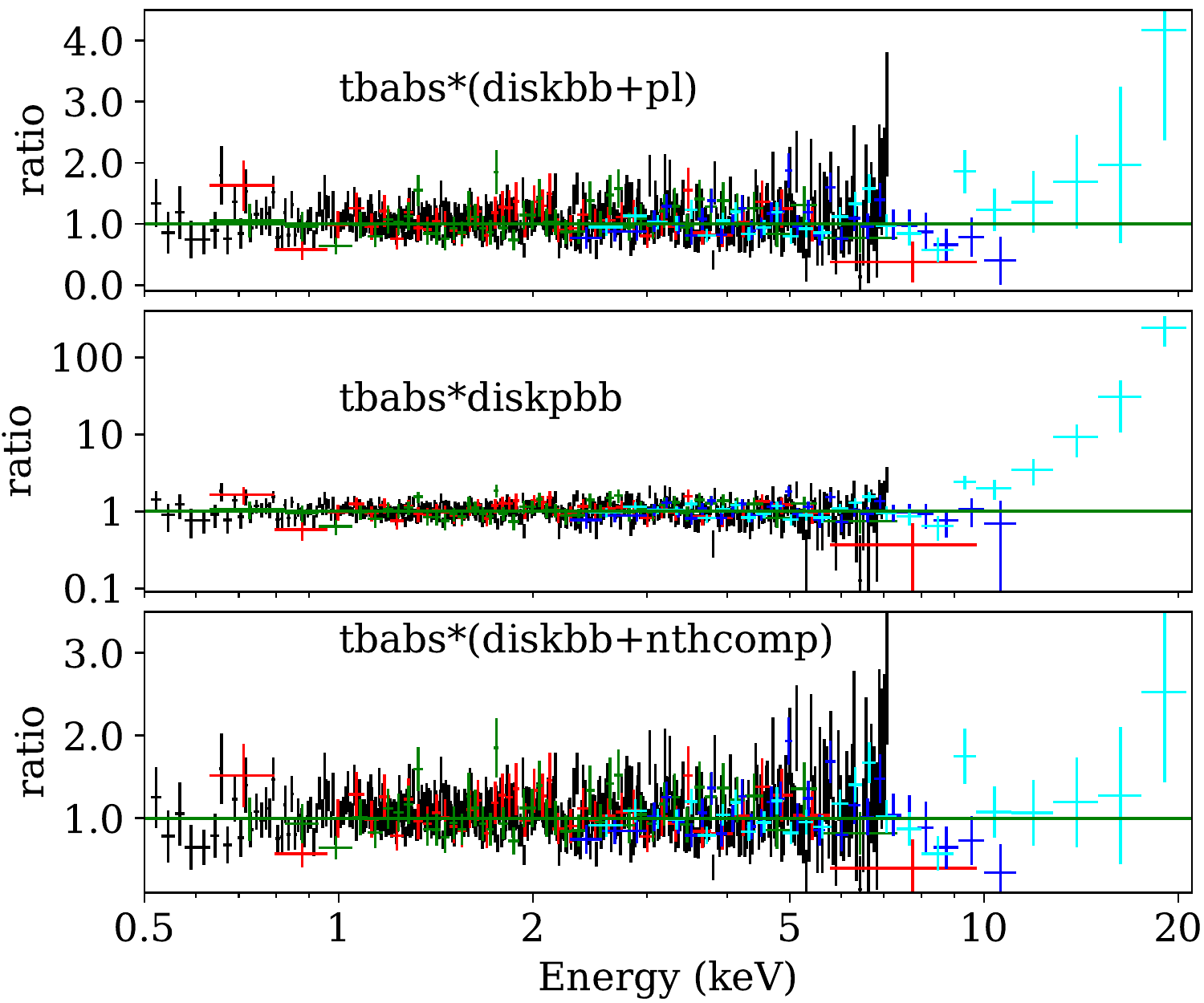}
    	\caption{The data to model ratio obtained from various model fits to the \emph{XMM-16} and \emph{NuSTAR-16} 
    	data set. Black, red, green, blue, and cyan denote data from EPIC-pn, MOS1, MOS2, FPMA and 
    	FMPB, respectively.}
    	\label{fig:xmm2_ratio}
	\end{figure}

During the 2016 observation, the emission above 10 keV was extremely weak, but the \emph{NuSTAR} spectrum 
shows a rollover above 10 keV. The 0.5--10 keV EPIC-pn mean count rate was 0.53 ct s$^{-1}$, whereas during 
2013 it was  1.25 ct s$^{-1}$.  Fig.~\ref{fig:xmm2_ratio} 
clearly demonstrates that the 0.5--20\,keV spectrum cannot be fitted by a single accretion disk model 
(neither \texttt{diskbb} nor \texttt{diskpbb}). Adding an extra \texttt{pl} or \texttt{nthcomp} 
component to  the \texttt{diskbb} model improves the fit significantly, but
 the ratio plot still shows a 
hard excess at higher energies. Furthermore, \cite{walton2018} 
compared the hard excess 
of the 2013 broadband 0.3--30 keV data set to the hard excess present in ULX pulsars M82 X-2, NGC 7793 P13 
and NGC 5907 ULX and conclude that the hard excess is consistent with being produced by an accretion column.

	\begin{figure}[]
    	\centering
    	\includegraphics[trim={0 0 0 0},width=0.48\textwidth]{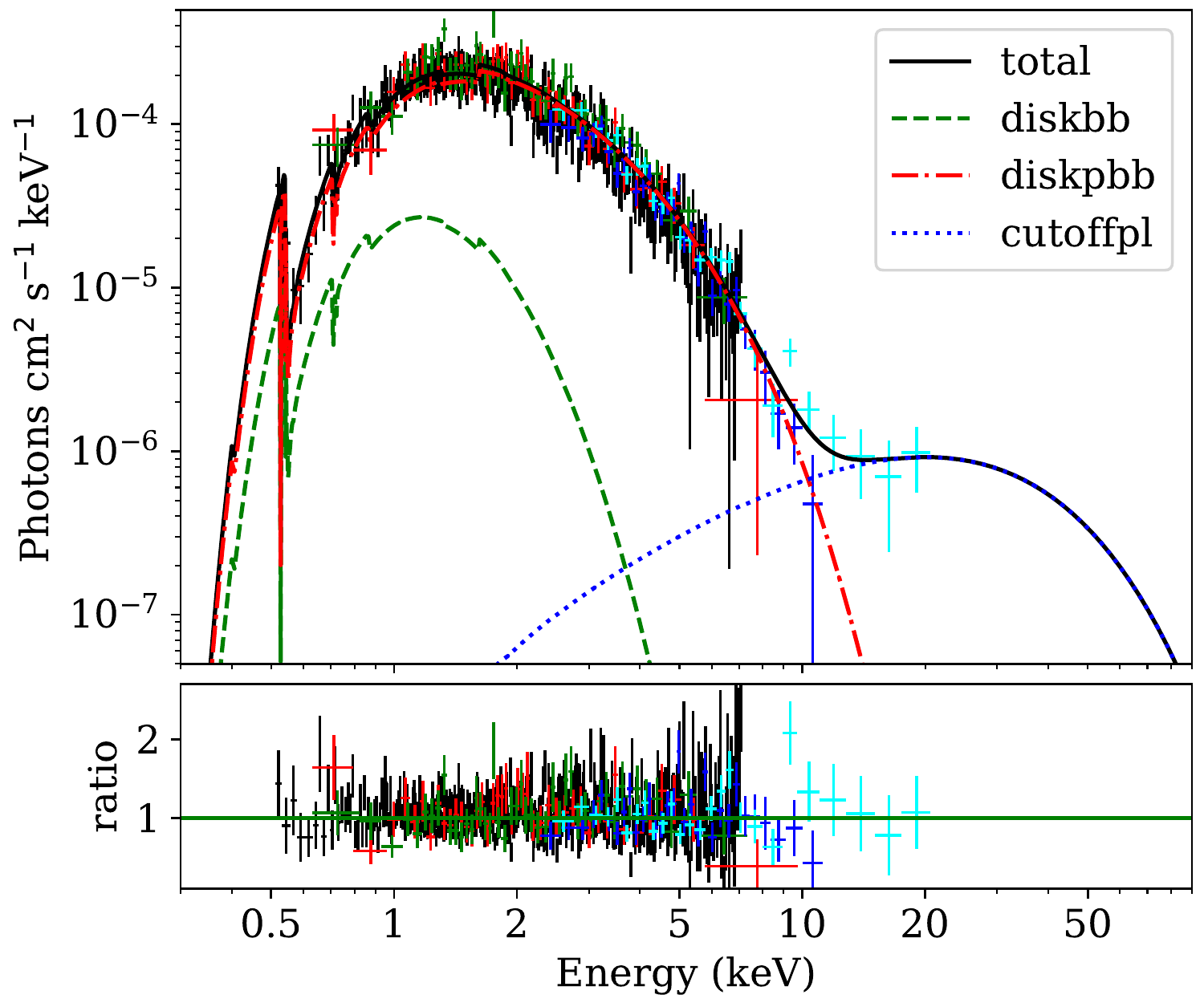}
    	\caption{Spectral fit to the \emph{XMM-16} and \emph{NuSTAR-16} data set. The total absorbed model is 
    	composed of \texttt{diskbb} for the outer sub-Eddington disk plus \texttt{diskpbb} for the inner super-Eddington 
    	disk and a cut-off power law (\texttt{cutoffpl}) for emission from the accretion column onto the neutron star. The color scheme 
   		of the data points is the same as in Fig.~\ref{fig:xmm2_ratio}.}
    	\label{fig:xmm4_ratio}
	\end{figure}

Therefore, we constructed a phenomenological accreting NS model similar to that of \cite{walton2018}, 
to fit the 2016 broadband 0.5-20 keV data. The model  is composed of a thin disk model, \texttt{diskbb}, 
for the outer sub-Eddington part of the disk up to the spherization radius $R_{\rm sph}$ plus the slim 
disk model \texttt{diskpbb} for the inner super-Eddington part from $R_{\rm sph}$ to the inner magnetospheric 
radius $R_{\rm m}$, and also includes a cut-off power-law model, \texttt{cutoffpl}, for emission from the accretion column. 
Due to the limited range of the broadband data, the energy cutoff is not well constrained;
therefore we freeze $E_{\rm cut}$ at 10 keV, since the ULX 
pulsars have observed cutoff energies around 5--15 keV 
\citep{brightman2016,israel2017b,walton2018a}.
We obtained an excellent fit, with $\chi^2$/$dof$=585/557 or $\chi^2_{\rm red}$=1.05. All fit parameters are 
given in Tab.~\ref{tab:acccol}
	
	\begin{table*}[]
	\tiny
    \centering
    \caption{Best parameters obtained from the fit of an accreting NS model to the \emph{XMM-16+NuSTAR-16} spectra. 
    The outer sub-Eddington disk is characterized by $T_{\rm in}$, while the inner super-Eddington accretion is characterized by $T_{\rm in,p}$. 
    The hard X-ray cutoff $E_{\rm cut}$ limits the accretion column temperature 
    and it is frozen during this fit. The value of reduced $\chi^2_{\rm red}=1.05$ for this fit.}
    \setlength{\tabcolsep}{2.6pt}               	
    \renewcommand{\arraystretch}{1.7}			
    \begin{tabular}{lrrccccccccc}
        \hline\hline
         & $N_{\rm H}$ & $kT_{\rm in}^{\rm bb}$ & $kT_{\rm in}^{\rm pbb}$&
         \multirow{2}{*}{$p$} & \multirow{2}{*}{$\Gamma$} &  $E_{\rm cut}$ &
           \multirow{2}{*}{$N_{\rm norm}^{\rm bb}$} & 
           $N_{\rm norm}^{\rm pbb}$ &
            $N_{\rm norm}^{\rm pl}$ &
         $F_{(0.5-10){\rm keV}}$ & $L_{(0.5-10){\rm keV}}$ \\
        \texttt{tbabs}*Model & $10^{21}$~[cm$^{-2}$] & [keV] &[keV]&& & [keV] &  & $\times10^{-2}$ & $\times10^{-8}$ & [erg s$^{-1}$ cm$^{-2}$]& [erg s$^{-1}$]  \\
        \hline
	\texttt{diskbb+diskpbb+cutoffpl}  &   $6.10^{+0.13}_{-0.12}$ & $0.31\pm0.01$
	&$1.33^{+0.06}_{-0.05}$ & $0.59^{+0.02}_{-0.01}$ & $-2.01^{+0.17}_{-0.12}$
	&  $10$ & $1.21\pm0.35$ & $2.14^{+0.60}_{-0.45}$ 
	& $1.40^{+0.51}_{-0.49}$ & $3.24\times10^{-12} $ & $ 6.84\times10^{39}$ \\
	\hline
    \end{tabular}
    \label{tab:acccol}
	\end{table*}  
	
The resultant fit with the unfolded spectrum is shown in Fig.~\ref{fig:xmm4_ratio}. The normalization 
of the accretion disk parameters gives radii of $R_{\rm sph}$=550 km and $R_{\rm m}$=73 km assuming 
the source was observed face-on and assuming the values of the color-correction factor mentioned previously. The 
spherization radius scales with mass accretion rate as $R_{\rm sph}$=$27\dot M_{\rm acc} R_{\rm s}/4\dot M_{\rm Edd}$ 
\citep{ss73}; equating this with the value obtained from the fit and considering a NS of 1.4\,M$_{\odot}$, 
the mass accretion rate is $\dot M_{\rm acc}\sim20\ \dot M_{\rm Edd}$. Similarly, the magnetospheric radius 
scales with the dipole magnetic field strength and the source luminosity as 
$R_{\rm m}$=$2.7\times10^8B_{12}^{4/7}M_{1.4 {\rm M}_{\odot}}^{1/7}L_{37}^{-2/7}$ cm where $B_{12}$, $L_{37}$ 
and $M_{1.4 {\rm M}_{\odot}}$ are the magnetic field strength in units of $10^{12}$ G, the bolometric 
luminosity in units of $10^{37}\ \rm erg\ s^{-1}$ and the mass of the NS in 1.4\,M$_{\odot}$ units, respectively 
\citep{lamb1973}. Considering the 0.5--10 keV luminosity as a proxy for the bolometric luminosity, we obtained a 
lower limit on the strength of the magnetic field is $>5\times10^{10}$ G.

\section{Discussion}
\label{sec:disc}

The combination of spectral fits and variability amplitudes of Circinus ULX5 supports 
state transitions in the source between hard/low UL state, represented in 
the \emph{XMM-01}, \emph{XMM-14}, and \emph{XMM-18F} data sets and
well fitted by a power-law model, while the
soft/high UL state, represented in \emph{XMM-13+NuSTAR-13} and
 \emph{XMM-18S}, and well fitted
      by a MCD + power-law model. 
      Furthermore, those states are separated by an intermediate UL state, 
      where the unabsorbed flux is only very roughly 50$\%$ higher than in the low state, but 
      the spectrum is already dominated by the disk component, \emph{Suz-H-06},
      \emph{Suz-L-06} and \emph{XMM-16+NuSTAR-16}. 
      Finally, the HID  displays two distinct ``islands'' ---  
      one for lower hardness ratio and lower luminosity and a second one for
      higher hardness and luminosity. 
      The transition track is not the same as observed in XRBs, since we do 
      not obtain typical ``q'' shape in the HID; rather, it looks more likely a HID diagram 
      for an atoll source, i.e. a NS with island and banana states \citep{altamirano2005,church2014}. 
      It is, of course, possible that the ULX may cover other regions of the HID simply not sampled by 
      the limited number of observations obtained so far. Nonetheless, we 
      have noticed significant changes in spectral and variability characteristics 
      between the ``islands'' on timescales of $\sim$7 months to several years. 
      Such clear state transitions have never reported in case of any other 
      individual ULX source. Our result strongly supports that Circinus ULX5
      switches between nearly- and super-Eddington modes 
      of accretion around a stellar-mass BH or possibly a NS as well. Note that XRBs display much faster state transitions, on the 
      order of days, and they can take typically months to trace out a complete evolution
      in the q-loop in the HID \citep[e.g., the persistent XRB system Cyg X-1, ][]{zhang1997}.
      The differences in timescales between ULX5 and XRBs may originate from the fact that transient 
      XRBs never become super-Eddingon, but we need more long term X-ray 
      monitoring of Circinus ULX5 to fully resolve this issue. 
      
	The short-timescale variability behavior also correlates with 
	 the spectral evolution.
	The fractional variability amplitudes in the 0.5--10 keV band lightcurves is large
	($ \sim$10--15\%) when the source is in hard/low state, and it is 
	small ($\sim$2\%)
	in the disk-dominated soft/high state. Galactic BH XRBs 
	show similar behavior,  with higher variability amplitudes 
	associated with coronal emission in the hard/low state
	 \citep{homan2001,churazov2001,munoz2011}. In general, ULXs do not show 
	short-timescale variability (both flux and spectral) and most 
	ULXs do not change over decades. 
	
	From our spectral fitting to multi-epoch data we obtained an inverse
	inner disk radius versus temperature relation, with a power of $-$1.90 from the \texttt{diskbb} 
	component normalization and $-$2.71 from the \texttt{diskpbb} component normalization. 
	In the $R_{\rm in}-T_{\rm in}$ diagram (Figs.~\ref{fig:Rin_Tin} and \ref{fig:pbbRin_Tin}), all inferred radii from disk-dominated spectra are near $6 R_{\rm G}$
	for a 10~M$_{\odot}$ BH, except for the cases
	where the inner disk temperatures and radii are not properly constrained. 
	The measured radii which are lying just below 6$R_{\rm G}$ could be increased in value, if Circinus ULX5 would be observed 
	at a higher inclination angle such as $\theta$>$70^{\circ}$, i.e.\ almost edge on. 
	It is very unlikely that Circinus ULX5 is observed
	at such a high inclination angle because in this scenario Circinus ULX5 would 
	appear as a soft ULX with $kT_{\rm in}\sim$0.1--0.3 keV, which most of the 
	emission from the inner disk 
	intercepted by a wind. However, a strong wind was not detected in the \textit{XMM-Newton} RGS data 
	(see Appendix~\ref{sec:rgs}). Circinus ULX5 has a
	rather high inner disk temperature, close to 2 keV, despite 
	its large luminosity. 
	
	Furthermore, the obtained disk flux and temperature relation 
	shows a positive correlation with high index $1.74^{+0.14}_{-0.11}$. 
	The best-fit value is not far from the typical $F_{\rm disk} \propto T_{\rm in}^2$
	relation expected from super-Eddington advection dominated accretion disk. Based on 
	this finding, together with the quite high value of $T_{\rm in}$ (in the range 
	of 1--2~keV; Fig~\ref{fig:diskflux_Tin}), it is most likely that the compact 
	object in Circinus ULX5 is a low mass BH (<10\,M$_{\odot}$), ruling out the hypothesis of IMBH.
	
	It is also possible that Circinus ULX5 can host a NS. 
	 We have demonstrated in this paper 
	that the \emph{XMM-16+NuSTAR-16} 0.5--20 keV 
	data can be well fitted by a three-component model composed of 
	MCD emission for the outer sub-Eddington disk, slim disk emission for the
	inner super-Eddington disk, and a cut-off power law responsible for the emission
	from the accretion column onto NS. From such a fit we can put rough constraints 
	on the mass accretion rate, $\sim$20 $\dot M_{\rm Edd}$, and on the strength of the 
	magnetic field, $>5\times 10^{10}$~G.
	In addition,
	 \cite{rozanska2018} obtained a very good fit by applying a 
	single model component composed of the emission from the NS surface plus accretion disk 
	emission to the broadband 0.3--30 keV 2013 data set.
	The fit is interesting since the distance to the source inferred from spectral fitting of 
	single model component agrees with other distant measurements
	(see Appendix~\ref{sec:dist} for discussion). 
	
\section{Conclusions}
\label{sec:concl}
We analyzed multi-epoch observations of Circinus ULX5 taken with \emph{Suzaku}, \emph{XMM-Newton} and \emph{NuSTAR} from 2001 to 2018. 
We performed spectroscopy and timing analyses to obtain the physical 
properties of the source and compare them with theoretical models. 
The source's unabsorbed 0.5--10 keV luminosity is extremely high: 
$\sim1.8\times10^{40}\ \rm erg\ s^{-1}$. The conclusions based on the 
timing analysis and various spectral fitting models are as follows:

\begin{enumerate}

\item The fitting of multi-epoch X-ray observations revealed three
  spectral states in which the source was observed: three observations
  caught the source in a hard/low UL, power law-dominated state; three
  observations caught the source in an intermediate state, with a
  disk-dominated spectrum but still relatively low flux; two
  observations caught the source in a soft/high disk-dominated state,
  where the flux had increased by a factor of four compared to the low/hard state.  The minimum
  observed timescale for a spectral state transition was $\sim7$
  months.

\item Circinus ULX5 exhibits short-timescale variability amplitudes that are
  correlated with spectral state.  The 0.5--10 keV fractional variability amplitude $F_{\rm var}$ is larger than 
  $ \sim$10-15\% in the hard/low state,  whereas in the thermal disk-dominated or high-flux state, variability is suppressed, and $F_{\rm var}$
  drops to $\sim$2\%.
  
\item The normalization of the MCD model indicates that the inner disk
  radius is equal or slightly smaller than the ISCO for a non-spinning 10\,M$_{\odot}$ BH. 
  This implies a stellar-mass compact object in Circinus ULX5. 
  As we do not know the spin of the compact object, there is a possibility that central 
  object can be highly spinning, but it is unlikely to be IMBH.

\item The flux and inner disk temperature follows a relation of
  $F_{\rm disk}\propto T_{\rm in}^{1.74^{+0.14}_{-0.11}}$ for the two
  component \texttt{diskbb} plus \texttt{diskpbb} model and the inner disk 
  radii obtained from the \texttt{diskpbb} normalization hints towards a 
  small mass compact objects. These results strongly suggests that 
  Circinus ULX5 accreting above the critical limit with a stellar mass compact 
  accretor.

\item The hard excess present in the broadband 0.5--20 keV
  \emph{XMM-16+NuSTAR-16} data can be interpreted as emission from the
  accretion column of a NS similar to that present in ULX pulsars. We
  constrain the mass accretion rate to be $\sim$20 $\dot M_{\rm Edd}$
  and the magnetic field strength to be $>5\times10^{10}$ G for a
  1.4\,M$_{\odot}$ NS, from fitting a phenomenological model of
  super-Eddington accretion.
\end{enumerate}

\begin{acknowledgements}
 AR was partially supported  by  Polish National Science 
Center grants No. 2015/17/B/ST9/03422, 2015/18/M/ST9/00541.
BDM acknowledges support from Ramóny Cajal Fellowship RYC2018-025950-I. AGM was 
partially supported by Polish National Science Center grants 2016/23/B/ST9/03123 
and 2018/31/G/ST9/03224. The work has made use of publicly available data from HEASARC 
Online Service, XMM-Newton Science Analysis System (SAS) developed by European Space 
Agency (ESA), and NuSTAR Data Analysis Software (NUSTARDAS) jointly developed by 
the ASI Science Data Center (ASDC, Italy) and the California Institute of Technology (USA). \\ 
\emph{Software}: Python \citep{python3}, Jupyter \citep{jupyter2016}, NumPy \citep{numpy2011,numpy2020}, matplotlib \citep{matplotlib2007}.
\end{acknowledgements}

%
%
\bibliographystyle{aa} 
\bibliography{refs}


\begin{appendix} 
\section{Distance to the source}
   \label{sec:dist}

We would like to pay attention that the commonly accepted distance to the 
Circinus galaxy is $4.2$\,Mpc given by NASA Extragalactic 
Database\footnote{https://ned.ipac.caltech.edu} (NED), derived with the use
of the Tully-Fisher method \citep{1977-Tully}. This method aims at finding the systemic velocity of a given galaxy 
by measuring the relation between global galaxian H{\sc I} profile width versus absolute magnitude. 
 The measurements of the line fluxes and optical magnitudes span many years  and are collected in 
 Extragalactic Distance Databases\footnote{http://leda.univ-lyon1.fr and http://edd.ifa.hawaii.edu} (EDD) 
 \citep{tully1988,tully2009a,karachentsev2013}, but the final distance does not agree with 
  the distance derived by the Hubble relation with the use of the same radial velocities reported in the above 
  EDD. Taking the  value of  radial velocity of the Circinus galaxy with respect to the Local Group (LG)
to be 204\,km\,s$^{-1}$, the distance  derived from Hubble relation, $D= v_{\rm LG}/H_{0}$, is
$D=2.8$\,Mpc, with the value of the Hubble constant 
$H_{0}=72$\,km\,s$^{-1}$\,Mpc$^{-1}$  \citep{2004-karachentsev}.

Another  distant measurement  from the Hubble relation was provided by \citet{Koribalski2004}, who reported 
the systemic velocity of the Circinus\footnote{The whole 
data set is available under a different name for the galaxy: HIPASS\,J1413-65,
 on http://vizier.cfa.harvard.edu/viz-bin/VizieR.} galaxy to be $v_{\rm sys} =434$\,km\,s$^{-1}$.
Knowing this value, the radial velocity with respect to the LG can be computed from the relation 
$v_{\rm LG}=v_{\rm sys} +300 \sin(l) \cos(b)$, where $l$ and $b$ are galactic coordinates. 
In the case of the Circinus galaxy, $l=311^{\circ}.3226$ and $b= -3^{\circ}.8076$, yielding
$v_{\rm LG} =209$\,km\,s$^{-1}$, reported in the above database. 
Even accepting the recent value of $H_{0}=75$\,km\,s$^{-1}$\,Mpc$^{-1}$, we 
get $D=2.78$\,Mpc. 
This distance value is consistent with that obtained by \citet{rozanska2018},
$D=2.60^{+0.05}_{-0.03}$~Mpc, 
wherein a single emission component from a non-spherical system containing 
NS and accretion disk was fitted to the same observations 
as presented in \citet{walton13}. 

In this paper we have accepted the distance of 4.2\,Mpc to Circinus galaxy, bearing 
in mind that Tully-Fisher relation is not very precise. Nevertheless, we have checked that for lower values of
distance, all luminosities reported in Tab.~\ref{tab:obs} become a factor of two lower, but still 
keep ULX5 in the regime of ultraluminous sources. 

\section{High resolution RGS spectra}
\label{sec:rgs}
\begin{figure}[]
    	\includegraphics[trim={0 0 0 0},width=0.48\textwidth]{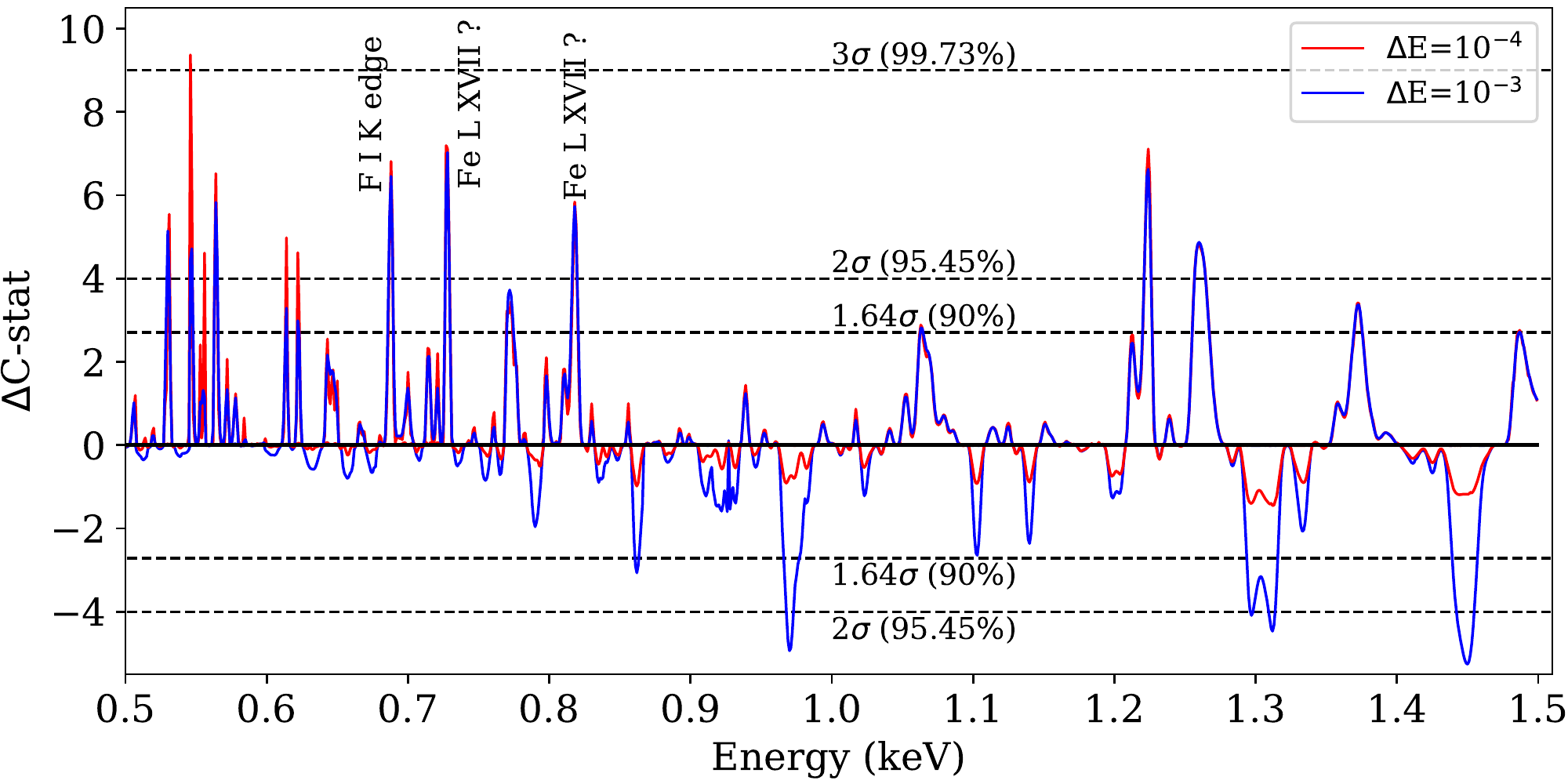}
    	\caption{The result of a line search from the first order RGS data by scanning the spectra  
          using a narrow Gaussian. Positive/negative values of $\Delta$$C$-stat denote
          candidate emission/absorption features, respectively. The F {\sc I} K edge effect is instrumental in
          nature. }
    	\label{fig:cstat_line}
	\end{figure}

	To search for evidence for disk-driven outflowing winds in Circinus ULX5, we analyzed spectra from the {\it XMM-Newton}
	reflecting grating spectrometers (RGS) to search for narrow emission and/or absorption lines from ionized species.
        For RGS analysis we utilized three long-exposure, on-axis observations of 
	Circinus ULX5: \emph{XMM-13}, \emph{XMM-14} and \emph{XMM-18S}. 
	To achieve a higher signal to noise ratio, we combined all first-order RGS1 and RGS2 spectra. 
	Even though the spectra looks very noisy, we 
	restrict ourselves to binning the data to a minimum of 3 counts per energy bin so that any possible the emission/absorption features do not smear out. As the 
	data is not of excellent quality, it is impossible to do a detailed modeling considering the optical depth, ionization parameter, wind column density, outflow 
	velocity, etc. However we can still search for the possible presence of line features by performing a simple Gaussian scan, focusing on the 0.5--1.5 keV 
	band, in which the RGS detector is the most sensitive. The fitting was done using Cash statistics \citep{cash1979} which utilizes the
        Poisson likelihood function, most 
	suitable for low numbers of counts per bin. The scan was done by adding a narrow Gaussian with a fixed energy and width (only the normalization was kept free) to the 
	continuum model, fitting, and checking if there is 
	improvement in the value of the $C$ statistic compared to the single power-law continuum model.
        We searched with a line centroid step size of 0.001 keV (1000 steps).
        For the width of the Gaussian line we 
	tested two values of $\sigma$: 1 and 0.1 eV. The improvement in the $C$ statistic as a function of energy is plotted in Fig.~\ref{fig:cstat_line} with positive/negative
        values of $\Delta$$C$-stat denoting  candidate emission/absorption-like features, respectively.
	The significance of the detection can directly obtained from the value of $\Delta$$C$-stat ($\Delta$$C$-stat=4 = 95.5\% confidence level), and there are multiple
        candidate features as shown in Fig.~\ref{fig:cstat_line}, most notably emission lines at 0.727 and 0.819 keV; the former is consistent with
        the rest-frame energy of Fe L {\sc xvii} 3s--2p, and the latter is in the vicinity of 0.826 keV,
        the rest-frame energy of Fe L {\sc xvii} 3d--2p. However this approach may overestimates the significance as it does not take into account the
        large number of trials. The actual significance should be estimated through Monte Carlo simulations. 
	We utilized the \texttt{simftest} tool available in {\sc xspec}; it generates simulated data sets based on the null hypothesis model
        (continuum emission only), and fits a Gaussian, yielding the value of $\Delta$$C$-stat associated with fitting photon
        noise. We performed 1000 trials for each individual candidate line identified with $\geq$ 2$\sigma$ confidence according to the $\Delta$$C$-stat. The 
	resulting significance values obtained via \texttt{simftest} are much lower compared to the $\Delta$$C$-stat test, and we cannot conclude that
        any of the candidate lines are real.
	The strongest visible lines achieve maximum significances of only 50.2\% 
	at 0.819 keV, 46\% for the line at 0.727 keV, and 
	42.1\% for the line at 0.532 keV.
	
\end{appendix}

\end{document}